\documentclass[twocolumn,aps,pra,showpacs, floatfix]{revtex4}
\usepackage{amsmath,amstext,amssymb,graphicx,bbm,color,calc,capt-of,ifthen}

\newcommand{\assign}{:=}
\newcommand{\emdash}{---}
\newcommand{\mathd}{\mathrm{d}}
\newcommand{\mathe}{\mathrm{e}}
\newcommand{\nocomma}{}
\newcommand{\tmem}[1]{{\em #1\/}}
\newcommand{\tmmathbf}[1]{\ensuremath{\boldsymbol{#1}}}
\newcommand{\tmop}[1]{\ensuremath{\operatorname{#1}}}

\newcommand{\tmtextit}[1]{{\itshape{#1}}}
\newcommand{\tmtextrm}[1]{{\rmfamily{#1}}}
\definecolor{grey}{rgb}{0.75,0.75,0.75}
\definecolor{orange}{rgb}{1.0,0.5,0.5}
\definecolor{brown}{rgb}{0.5,0.25,0.0}
\definecolor{pink}{rgb}{1.0,0.5,0.5}
\newcommand{\tmfloatcontents}{}
\newlength{\tmfloatwidth}
\newcommand{\tmfloat}[5]{
  \renewcommand{\tmfloatcontents}{#4}
  \setlength{\tmfloatwidth}{\widthof{\tmfloatcontents}+1in}
  \ifthenelse{\equal{#2}{small}}
    {\ifthenelse{\lengthtest{\tmfloatwidth > \linewidth}}
      {\setlength{\tmfloatwidth}{\linewidth}}{}}
    {\setlength{\tmfloatwidth}{\linewidth}}
  \begin{minipage}[#1]{\tmfloatwidth}
    \begin{center}
      \tmfloatcontents
      \captionof{#3}{#5}
    \end{center}
  \end{minipage}}

\newcommand{\Young}[6]{
\setlength{\unitlength}{#6pt}
\thinlines
\begin{picture}(#3,#1)(0,#5)
\multiput(0,0)(1,0){#4}{\line(0,1){#1}}
\multiput(0,0)(0,1){#2}{\line(1,0){#3}}
\end{picture}
}

\begin{document}

\title{Scavenging quantum information: Multiple observations of quantum
systems}\author{P.~Rap\v{c}an$^1$, J.~Calsamiglia$^2$, R.~Mu\~noz-Tapia$^2$,
E.~Bagan$^{2,3,4}$ and V.~Bu\v{z}ek$^{1, 5}$}
\affiliation{$^1$Research Center for Quantum Information, Institute of Physics, Slovak Academy of Sciences, D\'{u}bravsk\'{a}
cesta 9, 845 11 Bratislava, Slovak Republic
\\
$^2$F\'{\i}sica Te\`orica: Informaci\'{o} i Fen\`{o}mens Qu\`{a}ntics, 
Edifici Cn, Univ. Aut\`onoma de Barcelona, 08193 Bellaterra (Barcelona), Spain
\\
$^3$Department of Physics, Hunter College of the City University of New York, 695 Park Avenue, New York, NY 10021, USA
\\
$^4$Physics Department, Brookhaven National Laboratory, Upton, NY 11973, USA
\\
$^5$Faculty of Informatics, Masaryk University, Botanick\'a 68a, 602 00 Brno, Czech Republic
}

\begin{abstract}
  Given an unknown state of a qudit that has already been measured optimally,
  can one still extract any information about the original unknown state?
  Clearly, after a maximally informative measurement, the state of the system
  `collapses' into a post-measurement state from which the {\tmem{same}}
  observer cannot obtain further information about the original state of the
  system. However, the system still encodes a significant amount of
  information  about the original preparation for a second
  observer who is unaware of the  actions of the first one. We
  study how a series of independent observers can obtain, or scavenge,
  information about the unknown state of a system (quantified by the fidelity)
  when they sequentially measure it. We give  closed-form
  expressions for the estimation fidelity, when one or several
  qudits are available to carry information about the single-qudit state, and
  study the `classical' limit when an arbitrarily large number
  of observers can obtain (nearly) complete information on the system. In
  addition to the case where all observers perform most informative
  measurements we study the scenario where a finite number of observers
  estimate the state with equal fidelity,regardless of their position in
  the measurement sequence; and the scenario where all observers use identical
  measurement apparata (up to a mutually unknown orientation) chosen
   so that a particular observer's estimation fidelity is
  maximized.
\end{abstract}
\maketitle

\section{Introduction}\label{sec:Introduction}

One of the major questions in the interpretation of quantum mechanics is to
assert the reality of the wave function. Despite the opinion galore, which
also includes rejecting the necessity of attributing reality to a quantum
state {\cite{QTNeedsNoInterpretation}}, there is a consensus in that all
information on the state of a system is contained in the wave function (in the
sense that it provides the right outcome probabilities for each conceivable
measurement on the system). Since all this information is not accessible by a
single measurement and, on top of that, quantum formalism only gives outcome
probabilities, the meaning of the wave function has been
traditionally associated to an \tmtextit{infinite} ensemble of identically
prepared quantum systems (something which cannot be taken literally, but only
as a conceptual notion). Ground-breaking experiments with individual quantum
systems (see e.g.
{\cite{Gleyzes:2007fk,RevModPhys.73.565,Riebe:2004uq,RevModPhys.75.281}}) and
the advent of quantum information technology have brought the focus to
individual systems, away from the infinite ensemble picture.

The inherent limitation of quantum measurements to obtain complete information
about a system is intimately linked to the disturbance they cause on the
state. Clearly, if a given measurement extracts the maximum information on the
state of a system, then the same observer cannot obtain additional information
by performing further measurements on the system. This almost tautological
statement has the staring consequence that quantum measurements, no matter how
cautelous they are, inevitably disturb the state of the system (and thereby
erase any information on the original state of the system as far
as the same observer is concerned). The question remains: What
happens if the second measurement is performed by a second observer who
independently aims at gaining information about the original state of the
system? Indeed, we will see that a second independent observer, who does not
know the precise actions nor measurement results of the previous observers,
can still obtain {\tmem{some}} information on the original state of the
system. In this article will study how does this information degrade through a
sequence of independent measurements performed by independent observers.

We extend the study to the case where several copies of the unknown state are
provided to the observers  and,  more generally, when several copies
of the system are used to collectively encode the unknown single-copy state in
more efficient ways {\footnote{A related, restricted, problem of
multiple observations of quantum clocks, i.e. of an evolving phase reference,
has been studied in Ref.~{\cite{PhysRevA.62.062309}}.}}. This will allow us
to give a new insight on a thorny problem in quantum mechanics, namely the so
called quantum to classical transition {\cite{zurek:36}}. The microscopic
world is governed by the rules of quantum mechanics, which often seem to be in
sharp contrast with the rules of classical physics that govern the macroscopic
world. Since we both observe the classical macroscopic world and believe the
quantum description is the  correct one, the classical properties
of systems have to appear within the quantum description in a consistent
fashion. How exactly, quantitatively, do these classical properties emerge?
Before attempting to answer this question it is important to recognize what
are the essential features that seem to be so different in the classical and
quantum worlds. The problem at hand sheds some light on one of these
differential aspects, which is the fragility of the information encoded in
quantum states versus the enduring nature of classical information. 
Indeed the information encoded in a classical system can be accessed by an unlimited
number of (careful) observers without degrading while quantum
mechanics allows to retrieve some amount of information but this degrades as
the number of observers increases. We show that the larger the number of
copies of a system the more observers can gain a \tmtextit{sizeable} amount of
information encoded in the original state. In other words, the information
encoded in large collections of quantum systems of the same type behaves
``classically'', in the sense that it is robust with respect to observations.

In order to make quantitative statements we need to, first, quantify the
observers' ability to gain information about the unknown state and, second,
make a judicious, precise definition of what we mean by ``independent''
observers. Following
Refs.~{\cite{PhysRevLett.74.1259,PhysRevLett.80.1571,PhysRevLett.83.432,PhysRevLett.85.5230}},
we will here use as a figure of merit the quantum fidelity between unknown
input (pure) quantum state $\psi = \left| \psi \rangle \langle \psi \right|$
and the estimate, or guess, $\psi' = \left| \psi' \rangle \langle \psi'
\right|$, that each observer arrives at, based on his measurement outcome: $f
(\psi, \psi') = | \left\langle \psi' | \psi \right\rangle |^2$. The $k$'th
observer's success in gaining knowledge about original system is given by the
mean of the fidelity, $F_k$, over all possible input states and guesses:
\begin{equation}
  \label{eq:MeanFidelity} F_k = \int f (\psi_k, \psi_0) \mathd p (\psi_k,
  \psi_0),
\end{equation}
where $\mathd p (\psi_k, \psi_0) = \mathd \psi_0 \mathd \psi_k \tilde{p}
(\psi_k | \psi_0)$ is the joint probability of observer $k$ obtaining the
guess $\psi_k$ and the original unknown state being $\psi_0$, with $\tilde{p}
(\psi_k | \psi_0)$ being the corresponding conditional probability density, where a uniform prior distribution
for $\psi_0$ is assumed. The integrals run
over all joint events, i.e. over the set of all possible pure states $\psi_0,
\psi_k \in \mathcal{S} (\mathcal{H})$.

We want to define a scenario in which, one after the other, each observer
gains access to the quantum system, but lacks ``any'' information regarding
the actions of the previous observers. In principle, if no further directives
are given, each observer will choose his measurement based on his own prior
knowledge about it, i.e., he may describe the system as the mixed state that
results from taking the original input state and performing an average over
all the actions that the previous observers might have conceivably done, which
includes, e.g., the obstructive action of resetting the state of the system to
a fixed state. We want to find the limits on how well can independent
observers recover, or scavenge, information from the very same system after
successive measurements. We therefore assume that each observer will be
careful, i.e. willing to facilitate the task to following observers insofar as
this does not conflict with his own priorities. Accordingly, we assume that
the observers are free to agree, in advance, on a protocol as long as it does
not involve exchanging any information that would allow them to establish a
common canonical basis, or reference frame, in which to represent their states
and actions -- sharing such information would allow all the observers to
perform the very same projective measurement and hence all of them would
obtain the same measurement outcomes and estimate the original state with
equal precision {\footnote{ One can further relax this condition
and allow for forward communication between measurements as long as this is
invariant under the choice of basis.}. Mathematically, imposing this
condition is equivalent to describing the actions of the observers in a fixed
basis and then averaging the result over all possible choices of basis. More
succinctly, if a given observer performs a quantum operation $\$ \left[
\hat{\rho} \right]$ over the system in a state $\hat{\rho}$, to the other
observers the state will effectively transform as $\$_s [ \hat{\rho}] = \int
\mathd U U\$[U^{\dagger} \hat{\rho} U] U^{\dagger}$, where $\mathd U$ is the
Haar measure of the unitary group acting on the system's Hilbert space.

In order to complete the framework required to present all the results in this
article, we still need to specify what is the figure of merit and what type of
information can the observers share when several ($N > 1$) copies of the
$d$-dimensional system are available. One option is to consider this
multi-partite system as a single system and accordingly use the fidelity
between the collective input and guess states as a figure of merit, and use
the unitaries over the $d^N$-dimensional Hilbert space in order to compute the
effective transformation $\$_s [\hat{\rho}^{(N)}]$. However, here we will
follow a different, more physically motivated, approach: Since the observers
are asked to retrieve information on the encoded single-copy state, we use the
fidelity between $d$-dimensional states, and we consider that the observers
agree on using the same (mutually unknown) local basis for each of the copies,
i.e. the actions of the other observers are known up to a \tmtextit{rigid}
unitary operation $U(g) = g^{\otimes N}$, $g \in S U (d)$, i.e.
the effective transformation now reads 
$\$_s [ \hat{\rho}^{(N)}]= \int \mathd g U(g) \$[U \left( g \right)^{\dagger} \hat{\rho}^{(N)} U \left(
g \right)] U \left( g \right)^{\dagger}$. This characterization of independent
observers is equivalent to that encountered in protocols like
\tmtextit{aligning reference frames}
{\cite{PhysRevLett.87.167901,PhysRevLett.87.257903}} where the different
parties that do not share a reference frame try to exchange some information
{\cite{RevModPhys.79.555}}.

To allow for a further generalization of this scenario we will introduce the
 concept of a preparer. The role of the preparer is to encode the
state of a single system $\psi_0 = g \left| \psi_{\tmop{ref}} \rangle \langle
\psi_{\tmop{ref}} \right| g^{\dagger}$, $\left| \psi_{\tmop{ref}}
\right\rangle \in \mathcal{H}_d$, $g \in \tmop{SU} (d)$ into a collective state
of the Hilbert space of $N$ \ by a rigid rotation $\Psi = g^{\otimes N} \left|
\Psi_{\tmop{ref}} \rangle \langle \Psi_{\tmop{ref}} \right| g^{\dagger \otimes
N}$, $\left| \Psi_{\tmop{ref}} \right\rangle \in \mathcal{H}_D$, $D = d^N$,
i.e. the signal state $\Psi$ need not be one consisting of $N$ copies of the
initial state $\psi_0$, but the encoding $\varrho$ is still covariant with
respect to $S U (d)$ and its tensor-product representation --- covariant, for
short. Considering covariant encodings only is actually not a restriction --
as we will argue in Section~\ref{sec:greedy} this follows from the
quantum-operations averaging discussed above.

The paper is structured as follows: In Section~\ref{sec:General} we discuss
general considerations relevant for the rest of the paper. In
Section~\ref{sec:greedy} we analyze the essential scavenging setting in which
each observer maximizes the quality of its own estimate. We call this scenario
the  ``greedy'' strategy. We study the cases of: i) the optimal
general encoding and ii) a symmetric encoding of the qudit state into a signal
state consisting of $N$ copies of the encoded state.
In Section~\ref{sec:WeakMeasurements} we study the action of repeated 
weak measurements on a state.
Sections~\ref{sec:FairObservers} and \ref{sec:FavorLast} are devoted to generalizations of the basic
setting in which the information about the encoded state is redistributed
among the observers by making use of weak measurements. We first consider what
we call the {\tmem{egalitarian}} strategy, where the measurement apparata are
chosen so as to provide the same quality of the estimation for all observers.
We then study a complementary scenario, where all observers use the very same
apparatus, but tailored in such a way that a {\tmem{privileged}} observer
obtains the best estimate. We present our conclusions in
Section~\ref{sec:Conclusion}. Three appendices follow with some technical
details used in the main text.

\section{Scavenging qudit information: General
considerations}\label{sec:General}

The computation of the fidelity, Eq.~(\ref{eq:MeanFidelity}), requires the
evaluation of the $k$th observer's conditional probability density $\tilde{p}
(\psi_k | \psi_0)$ of obtaining a guess $\psi_k$ given the state to estimate
$\psi_0$. Naturally, this quantity depends on the preparation (the way
$\psi_0$ is encoded into a signal state), on the $k$th observer's measurement
and the guessing strategies, and on whatever happened in-between. We may decompose
each particular conditional joint event into a sum over histories ---
intermediate events such as measurements apparata choices, obtained
measurement outcomes, or guesses made based on the outcomes --- which may have
led to the event $(\psi_k | \psi_0)$. It is clear that even though
probabilistic strategies in the encodings and measurements choices are
possible, these perform equally well as deterministic strategies which are
given by averaging, with their respective probabilities, over those
strategies.

Without loss of generality, we may write
\begin{eqnarray}
  \tilde{p} (\psi_k | \psi_0) & = & \tmop{Tr} \left[
  \tilde{\mathcal{M}}^{(k)}_{\psi_k} \chi_{k - 1} \circ \ldots \circ \chi_1
  \left( \varrho_0 (\psi_0) \right) \right], 
  \label{eq:p(k|0)ChannelApproachEverythingCovariant}
\end{eqnarray}
where $\varrho_0 (\psi_0)$ is the state provided by the preparer, $\chi_i$ is
a channel induced by an averaged (over all unknowns) measurement of the $i$th
observer and $\tilde{\mathcal{M}}^{(k)}$ is the operator density, with respect
to the measure $\mathd \psi_k$, of the POVM $\mathcal{M}^{(k)}$ performed by
the $k$th observer, with measurement outcomes labeled by the guesses $\psi_k$
the observer makes. Note that should several outcomes lead to the same guess,
$\psi_k$, the POVM element corresponding to $\psi_k$ is given by sum of all
effects (i.e. POVM elements) of such outcomes. 

Eq.~(\ref{eq:p(k|0)ChannelApproachEverythingCovariant}) can also be written as
a decomposition over all intermediate observers' guesses, $\psi_i$,
\begin{eqnarray}
  \tilde{p} (\psi_k | \psi_0) & = & \int \mathd \psi_{k - 1} \ldots \int
  \mathd \psi_1 \tmop{Tr} \left[ \tilde{\mathcal{M}}^{(k)}_{\psi_k} \right.
  \nonumber\\
  &  & \times \left( \tilde{\mathcal{I}}^{(k - 1)}_{\psi_{k - 1}} \circ
  \ldots \right. \left. \left. \circ \tilde{\mathcal{I}}^{(1)}_{\psi_1}
  \right) \left( \varrho_0 (\psi_0) \right) \right], 
  \label{eq:p(k|0)EverythingCovariant}
\end{eqnarray}
where $\tilde{\mathcal{I}}^{(i)}$ is the density of the $i$th observer's
average (over all unknowns) quantum instrument $\mathcal{I}^{\left( i
\right)}$ with measurement outcomes labeled by the guesses $\psi_i$. Quantum
instruments, or instruments for short, introduced by Davies and Lewis
{\cite{springerlink:10.1007/BF01647093}}, are the standard mathematical tool
used to describe the measurement process when one is interested not only in
probabilities of measurement outcomes, but also in the post measurement state.
An instrument -- or an operation-valued measure -- assigns to a set
$\mathcal{B}$ of measurement outcomes an operation $\mathcal{I}_{\mathcal{B}}$
that provides a transformation rule of the state due to the measurement as
well as the probability of the outcome, which is given by the trace of the
transformed (post-measurement) state.

In our case, due to the limited mutual knowledge of the observer's (and
preparer's) actions, the average instruments $\mathcal{I}^{\left( i \right)}$
are covariant with respect to $S U \left( d \right)$ and its symmetric
representation $U$, i.e. $\forall \hat{\rho} \in \mathcal{S} \left(
\mathcal{H}_D \right) \nocomma, \forall g \in S U \left( d \right)$,
\begin{eqnarray}
  & \tilde{\mathcal{I}}^{\left( i \right)}_{g \psi g^{\dagger}} ( \hat{\rho})
  = U \left( g \right) \tilde{\mathcal{I}}^{\left( i \right)}_{\psi} (U \left(
  g \right)^{\dagger} \hat{\rho} U \left( g \right)) U \left( g
  \right)^{\dagger} . & 
\end{eqnarray}
For the same reason the (averaged) encoding $\varrho_0$, is covariant with
respect to $S U \left( d \right)$ and its symmetric representation $U$, i.e.
$\forall \psi \in \mathcal{S} \left( \mathcal{H}_d \right) \nocomma, \forall g
\in S U \left( d \right)$,
\begin{eqnarray}
  & \varrho_0 (g \psi g^{\dagger}) = U \left( g \right) \varrho_0 (\psi) U
  \left( g \right)^{\dagger} . & 
\end{eqnarray}
Obviously, the channels in
Eq.~(\ref{eq:p(k|0)ChannelApproachEverythingCovariant}), which are induced by
the instruments in Eq.~(\ref{eq:p(k|0)EverythingCovariant}), \ are \ also
covariant with respect to $U$, i.e. $\forall \hat{\rho} \in \mathcal{S} \left(
\mathcal{H}_D \right) \nocomma \nocomma, \forall g \in S U \left( d \right)$,
\begin{eqnarray}
  & \nocomma \chi_i (U \left( g \right) \hat{\rho} U \left( g
  \right)^{\dagger}) = U \left( g \right) \chi_i ( \hat{\rho}) U \left( g
  \right)^{\dagger} . & 
\end{eqnarray}
Subsequently, the average ``encoding'' $\chi_{k - 1} \circ \ldots \circ \chi_1
\left( \varrho_0 \left( \cdot \right) \right)$ is also is covariant with
respect to $S U \left( d \right)$ and its symmetric representation $U$. The
estimation of $\psi_0$ can be viewed as the estimation of $\chi_{k - 1} \circ
\ldots \circ \chi_1 \left( \varrho_0 \left( \psi_0 \right) \right)$ with
$\varrho_0 \left( \psi_0 \right) $ known to be restricted to the invariant
family
\begin{equation}
  \label{eq:U-invariantFamily} \{\varrho_0 (\psi_0) = U \left( g \right)
  \varrho_0 (\psi_{\tmop{ref}}) U \left( g \right)^{\dagger}, g \in S U (d), U
  \left( g \right) = g^{\otimes N} \}
\end{equation}
distributed according to the Haar measure $\mathd \mu (g) = \mathd \psi_0$.
In other words, we have a covariant optimal estimation problem,
which, with the fidelity as the cost function, can always be solved by a
covariant POVM {\cite{Holevo1982Prob.andStat.Asp.of.Q.Theory}}. Hence, without
loss of generality, we may assume the POVM $\mathcal{M}^{(k)}$ to be covariant
with respect to $S U \left( d \right)$ and its symmetric representation $U$,
i.e. $\forall \psi \in \mathcal{S} \left( \mathcal{H}_d \right), \forall g \in
S U \left( d \right)$,
\begin{eqnarray}
  & \tilde{\mathcal{M}}^{\left( k \right)} (g \psi g^{\dagger}) = U \left( g
  \right) \tilde{\mathcal{M}}^{\left( k \right)} (\psi) U \left( g
  \right)^{\dagger} . & 
\end{eqnarray}

Eqs.~(\ref{eq:p(k|0)ChannelApproachEverythingCovariant}) and
(\ref{eq:p(k|0)EverythingCovariant}) tell us how to proceed further. We search
for pairs of covariant encodings $\varrho_0$ and POVM(s) $\mathcal{M}^{(1)}$
fulfilling a desired property of the average fidelity $F_1$ (e.g. maximizing
$F_1$, possibly given additional constraints) for the invariant family of
states Eq.~(\ref{eq:U-invariantFamily}). Having at least one such pair
$\{\varrho_0, \mathcal{M}^{\left( 1 \right)} \}$, one can evaluate $F_1$.
Next, for {\tmem{each}} possible POVM $\mathcal{M}^{(1)}$,
obtained in the previous step, consider all covariant quantum instruments
$\mathcal{I}^{(1)}$ compatible with the POVM, and calculate the set of
covariant channels $\chi_1$ which are induced by any of those instruments.
Next, search for covariant POVM(s) $\mathcal{M}^{(2)}$ fulfilling a desired
property of the average fidelity $F_2$ for the average states from the
covariant families $\{\chi_1 (\varrho_0 (\psi_0)) = U \left( g \right) \chi_1
(\varrho_0 (\psi_{\tmop{ref}})) U \left( g \right)^{\dagger}, g \in S U (d), U
\left( g \right) = g^{\otimes N} \}$, distributed as governed by the Haar
measure $\mathd \mu (g)$, given by the actions of all channel(s) $\chi_1$ from
the previous step, and so on.

The task is greatly simplified by the possibility to restrict oneself to
covariant apparata and channels. If the optimal covariant apparata turn out to
be unique at each step, the task becomes even simpler. But, even in this case, 
one has still to calculate the induced channel at each step to
obtain the set of average states for the next optimization.

In the following sections we will study various scenarios which we separate in
two groups: those which require maximally informative measurements, where the
action for each observer can be interpreted as a measure and prepare channel;
and those where the conditions of the problem require that the observers
perform weak measurements extracting less information from the system.

\section{Greedy strategy}\label{sec:greedy}

Let us now specialize to the case of 'greedy' observers who primarily want to
maximize the fidelity of their own guesses. This is precisely the main
scenario that motivates this work  {\footnote{ Preliminary partial
results concerning the greedy scenario have been reported in the proceedings
{\cite{Phys.Scr.2010014059}}.}. 
Here the task at hand can be reduced
to the problem of a preparer and single-observer encoding/estimation of
quantum states embedded in larger systems. Solutions to the latter problem are
often known
({\cite{Holevo1982Prob.andStat.Asp.of.Q.Theory,PhysRevLett.80.1571,PhysRevLett.74.1259,PhysRevA.63.052309,PhysRevLett.81.1351,PhysRevA.61.022113}}).

In this scenario, each observer performs the best estimation he can ---in
other words, there exists no additional measurement he could perform that
would increase the fidelity of his guess, which was obtained based on his
original measurement. It follows that the post-measurement state after the
$i$th measurement can depend on the original state $\psi_0$ only indirectly,
through the obtained guess $\psi_i$, hence the corresponding instrument can be viewed as a measure-and-prepare channel.
Thus, we can rewrite the
Eq.~(\ref{eq:p(k|0)EverythingCovariant}) in a factorized form
\begin{eqnarray}
  \tilde{p} (\psi_k | \psi_0) & = & \int \mathd \psi_{k - 1}  \tilde{p}
  (\psi_k | \psi_{k - 1}) \ldots \nonumber\\
  &  & \ldots \int \mathd \psi_0  \tilde{p} (\psi_1 | \psi_0), 
  \label{eq:p(k|0)IntoHistoriesFullDisturb}
\end{eqnarray}
with
\begin{eqnarray}
  \tilde{p} (\psi_i | \psi_{i - 1}) & = & \tmop{Tr} [
  \tilde{\mathcal{M}}^{(i)} (\psi_i) \varrho_{i - 1} (\psi_{i - 1})], 
\end{eqnarray}
where $\varrho_{i - 1} (\psi_{i - 1})$ is the post-measurement state after the
$\left( i - 1 \right)$th measurement given the obtained guess has been
$\psi_{i - 1}$.

Using the Bloch-vector formalism we may rewrite the average fidelity
Eq.~(\ref{eq:MeanFidelity}) as
\begin{eqnarray}
  F^{(k)} & = & \frac{1}{d} \left[ \right. 1 + (d - 1) \nonumber\\
  &  & \times \int \mathd \psi_0 \mathd \psi_k \tmmathbf{n} (\psi_k) \cdot
  \tmmathbf{n} (\psi_0) \tilde{p} (\psi_k | \psi_0) \left. \right] 
  \label{eq:AvgFidelityInBlochVectorLanguage}
\end{eqnarray}
where $\tmmathbf{n} (\psi)$ stands for the generalized Bloch vector of a pure
state $\psi \in \mathcal{S} (\mathcal{H}_d)$ (see
Appendix~\ref{sec:BlochVectorEvolution} for details).

As we argued in Section~\ref{sec:General}, the average instrument
$\mathcal{I}^{\left( i \right)}$ and thus the induced POVM
$\mathcal{M}^{\left( i \right)}$ and the encoding $\varrho_i$, $i < k$, are
covariant, while the POVM $\mathcal{M}^{\left( k \right)}$ can be chosen to be
covariant. Hence, without loss of generality, we may restrict our attention
the optimal covariant POVMs for the set of equiprobable states of the
invariant family $\{\varrho_{i - 1} (\psi_{i - 1}) = U \left( g \right)
\varrho_{i - 1} (\psi_{\tmop{ref}}) U \left( g \right)^{\dagger}, \psi_{i - 1}
= g \psi_{\tmop{ref}} g^{\dagger}, U \left( g \right) = g^{\otimes N}, g \in S
U \left( d \right) \}$.

For such a situation, we show in Appendix~\ref{sec:BlochVectorEvolution} that
\begin{equation}
  \label{eq:BlochVectorEvolution} \int \mathd \psi_{i - 1}  \tmmathbf{n}
  (\psi_{i - 1}) \tilde{p} (\psi_i | \psi_{i - 1}) = \Delta_i \tmmathbf{n}
  (\psi_i),
\end{equation}
where $\Delta_i$ is a number. In other words, after the measurement of the
$i$th observer, the output state will be characterized by a ``shrinked''
version of the input Bloch-vector.

Plugging Eqs.~(\ref{eq:p(k|0)IntoHistoriesFullDisturb}) and
(\ref{eq:BlochVectorEvolution}) into
Eq.~(\ref{eq:AvgFidelityInBlochVectorLanguage}) we have
\begin{eqnarray}
  F^{(k)} & = & \frac{1}{d} \left( 1 + (d - 1) \prod_{i = 1}^k \Delta_i
  \int_{\mathbbm{S}} \mathd \psi_k \tmmathbf{n} (\psi_k) \cdot \tmmathbf{n}
  (\psi_k) \right) \nonumber\\
  & = & \frac{1}{d} \left( 1 + (d - 1) \prod_{i = 1}^k \Delta_i \right) . 
  \label{eq:FGeneral}
\end{eqnarray}
Thus, successive maximizations of $F_1, F_2, \ldots, F_k$ are achieved via
successive maximizations of $\Delta_1, \ldots, \Delta_k$. The maximization of
$\Delta_i$ is over the pair -- covariant encoding $\varrho_{i - 1}$, covariant
POVM $\mathcal{M}_i$ \ for the set of states $\left\{ \varrho_{i - 1} (\psi_{i
- 1}) \right\}$ with unknown, hence equiprobable, previous observer's guess
$\psi_{i - 1} = \left| \psi_{i - 1} \rangle \langle \psi_{i - 1} \right| \in
\mathcal{S} (\mathcal{H}_d)$.

If the initial encoding $\varrho_0$ has been optimal, then one cannot achieve
a better performance than if we take $\forall i, \varrho_i \equiv \varrho_0$,
i.e. $\Delta_i = \max \Delta_1 = : \Delta$. Hence, the maximum $\mathcal{F}_k$
of the average fidelity $F_k$ if all $F_i$, $i < k$ are, one-after-another,
maximal, reads
\begin{equation}
  \label{eq:FGeneralSameProbabilities} \mathcal{F}_k = \frac{1}{d} \left[ 1 +
  (d - 1) \Delta^k \right].
\end{equation}
The situation is different if the initial encoding has been restricted by some
additional requirements, e.g. encoding into copies of the state $\psi_0$,
which turns out to be a sub-optimal encoding with $\Delta_{\tmop{sym}}$. Then,
for $i \geq 1$, the best strategy is, naturally, to take $\varrho_i$ equal to
the unrestricted optimal $\varrho_0$. That is, for the problem of $k$ greedy
observers independently estimating $N$ copies of an unknown state the fidelity
will read
\begin{equation}
  \text{$\label{eq:FGeneralSameProbabilitiesFirstSym} \mathcal{F}_k =
  \frac{1}{d} \left[ 1 + (d - 1) \Delta_{\tmop{sym}} \Delta^{k - 1} \right]
  .$}
\end{equation}

In the following subsections we give the explicit expressions \ for the
fidelity $\mathcal{F}_k$ for some relevant cases, which amounts to solving the
relatively straightforward the single-observer problem (i.e. computing
$\Delta$).

\subsection{The Fidelity for the optimal $N$-qubit
encoding}\label{sec:GeneralQubits}

Here, we treat the optimization of the qubit state encoding with full
generality. The optimal encoding of a single qubit state (or, equivalently, of
a \ spatial direction corresponding to a spin-1/2 particle) into $U$-invariant
family of \ $N$-qubit states is \ given by Bagan \tmtextit{et al.} in
Ref.~\,{\cite{PhysRevA.63.052309}}.

The optimal signal state ($k = 0$) as well as the state prepared after the
$k$th measurement, $k > 1$, reads
\begin{equation}
  \label{eq:OptState4Qubits} \varrho_k ( \tmmathbf{m}_k) = U ( \tmmathbf{m}_k)
  \hspace{0.25em} |A \rangle \langle A| \hspace{0.25em} U^{\dagger} (
  \tmmathbf{m}_k) ; \hspace{1em} k \geq 0,
\end{equation}
where
\[ |A \rangle = \sum_{j = 0}^{N / 2} A_j |j, 0 \rangle, \]
(for simplicity we assume that $N$ is even) with the coefficients $A_j$ such
that $|A \rangle$ is the eigenvector corresponding to the maximal eigenvalue
of the matrix
\begin{equation}
  \label{eq:3DiagMatrix} \left(\begin{array}{ccccc}
    0 & c_{l - 1} & 0 & \ldots & 0\\
    c_{l - 1} & \ddots & \ddots & \ddots & \vdots\\
    0 & \ddots & \ddots & c_2 & 0\\
    \vdots & \ddots & c_2 & 0 & c_1\\
    0 & \cdots & 0 & c_1 & 0
  \end{array}\right),
\end{equation}
where
\begin{equation}
  l = \frac{N}{2} + 1
\end{equation}
and
\begin{eqnarray}
  c_i & = & \frac{i}{\sqrt{(2 i + 1) (2 i - 1)}} . 
\end{eqnarray}
The operator density of the corresponding optimal measurement can also be
found in {\cite{PhysRevA.63.052309}}:
\begin{equation}
  \label{eq:OptPOVM4Qubits} M^{(k)} ( \tmmathbf{m}_k) = U ( \tmmathbf{m}_k)
  \hspace{0.25em} |B \rangle \langle B| \hspace{0.25em} U^{\dagger} (
  \tmmathbf{m}_k) ; \hspace{1em} k > 1,
\end{equation}
where
\begin{equation}
  |B \rangle = \sum_{j = 0}^{N / 2} \sqrt{2 j + 1} |j, 0 \rangle .
\end{equation}
In this case
\begin{equation}
  \label{eq:QubitsDeltaOptEven} \Delta = x_{N / 2 + 1},
\end{equation}
where $x_{N / 2 + 1}$ is the largest zero of the Legendre polynomial $P_{N / 2
+ 1} (x)$. Thus
\begin{equation}
  \mathcal{F}^{\text{\tmtextrm{op}}}_k = \frac{1}{2} \left[ 1 + x_{N / 2 +
  1}^k \right] .
\end{equation}
Asymptotically, it is known that
\begin{equation}
  x_n = 1 - \frac{\xi_0^2}{2 n^2} + \cdots,
\end{equation}
where $\xi_0 = 2.4$ is the first zero of the Bessel function~$J_0 (x)$. Hence,
asymptotically (for $N \rightarrow \infty$),
\begin{equation}
  \Delta \cong 1 - \frac{2 \xi_0^2}{N^2},
\end{equation}
and
\begin{equation}
  \mathcal{F}^{\text{\tmtextrm{op}}}_k \cong \frac{1}{2} \left[ 1 + \left( 1 -
  \frac{2 \xi_0^2}{N^2} \right)^k \right] .
\end{equation}

In the case where the signal state is given as $N$ copies of the unknown
state, i.e. $\psi_0^{\otimes N}$, \ the first ($k = 1$) shrinking factor needs
to be replaced by that of the well known $N$-qubit pure state estimation
$\Delta_{\tmop{sym}} = \frac{N}{N + 2}$ (see next subsection for general
qu$d$it derivation). So that,
\begin{eqnarray}
  \mathcal{F}^{N \tmop{copy}}_k & = & \frac{1}{2} \left[ 1 + \frac{N}{N + 2}
  x_{N / 2 + 1}^{k - 1} \right] \\
  & \cong &  \frac{1}{2} \left[ 1 + \left( 1 - \frac{2}{N^{}} \right) \left(
  1 - \frac{2 \xi_0^2}{N^2} \right)^{k - 1} \right] . 
\end{eqnarray}
We note that by allowing operations to act on the whole Hilbert space of $N$
qubits provides a significant advantage with respect to strategies relying on
the encoding into $N$ copies, $\psi_0^{\otimes N}$, (which lies the in
completely symmetric \ subspace): \ in the latter case the maximum fidelity is
approached as $1 / N$, in contrast to the $1 / N^2$ behaviour found in the
optimal case {\emdash} see end of this Section for a more detailed discussion.

\subsection{The Fidelity for $N$ parallel qu$d$its}\label{sec:ParallelQudits}

For general $d$-dimensional states the general optimization is a hard problem
to solve. In this section we will limit ourselves to the situation where the
measurement apparata of the different observers are restricted to operate in
the Hilbert space of the initial state, which we also take to be $N$ copies of
an arbitrary pure qu$d$it state. \ That is, during the whole measurement
sequence the system will be constrained in the \text{totally symmetric
subspace of the state space $\mathcal{S} (\mathcal{H}_D)$}. A natural way to
impose this limitation could be to require that if in the fortunate, but
`extremely rare', event that an observer guesses the input state correctly,
then the output state should be left in exactly the same collective state as
the input.{\color{blue} }

The POVM $\mathcal{M}=\mathcal{M}^{\tmop{sym}}$ optimal for the encoding into
copies is known to be the extremal covariant POVM
{\cite{Holevo1982Prob.andStat.Asp.of.Q.Theory}} with the operator density on
the relevant, symmetric, subspace given by
\begin{equation}
  \label{eq:CovMeasNCopies} \tilde{\mathcal{M}}^{\tmop{sym}} (\psi) =
  d^{\text{\tmtextrm{sym}}}_N \left| \psi \rangle \langle \psi
  \right|^{\otimes N}
\end{equation}
where
\begin{eqnarray}
  \left| \psi \right\rangle^{\otimes N} = (g \left| \psi_{\tmop{ref}}
  \right\rangle)^{\otimes N}, & g \in S U (d), & \left| \psi_{\tmop{ref}}
  \right\rangle \in \mathcal{H}_d .  \label{eq:CovariantFamilyNQuditCopies}
\end{eqnarray}

The maximal single-observation fidelity is
\begin{eqnarray}
  \mathcal{F}^{\tmop{sym}}_1 & = & \int \mathd \psi \mathd \hat{\psi}
  \hspace{0.25em} | \langle \psi | \hat{\psi} \rangle^N |^2 p ( \hat{\psi} |
  \psi) \nonumber\\
  & = & d^{\text{\tmtextrm{sym}}}_N \int \mathd \psi \hspace{0.25em} |
  \langle \psi | \psi_{\tmop{ref}} \rangle |^{2 (N + 1)} \nonumber\\
  & = & d^{\text{\tmtextrm{sym}}}_N \langle
  \Psi^{\text{\tmtextrm{sym}}}_{\tmop{ref}} | \left[ \int \mathd \mu (g)
  \hspace{0.25em} \mathcal{U}(g) \right. \nonumber\\
  &  & \times | \Psi_{\tmop{ref}}^{\text{\tmtextrm{sym}}} \rangle
  \left\langle \Psi_{\tmop{ref}}^{\text{\tmtextrm{sym}}}
  |\mathcal{U}(g)^{\dagger} \right] |
  \Psi^{\text{\tmtextrm{sym}}}_{\tmop{ref}} \rangle \nonumber\\
  & = & \frac{d^{\text{\tmtextrm{sym}}}_N}{d_{N + 1}^{\tmop{sym}}}, 
  \label{eq:F1Parallel}
\end{eqnarray}
where $| \Psi_{\tmop{ref}}^{\text{\tmtextrm{sym}}} \rangle = |
\psi_{\tmop{ref}} \rangle^{\otimes (N + 1)}$ , and the dimension of the
completely symmetric representation is given by
\begin{equation}
  \label{eq:dSymN} d^{\text{\tmtextrm{sym}}}_N = \binom{N + d - 1}{N} .
\end{equation}
Substituting Eq.~(\ref{eq:dSymN}) into Eq.~(\ref{eq:F1Parallel}) we get
\begin{equation}
  \label{eq:FidSym1st} \mathcal{F}^{\tmop{sym}}_1 = \frac{N + 1}{N + d} .
\end{equation}
Using Eq.~(\ref{eq:FGeneralSameProbabilities}) we have
\begin{eqnarray}
  \mathcal{F}^{\tmop{sym}}_k & = & \frac{1}{d} \left[ 1 + (d - 1) \left(
  \frac{N}{N + d} \right)^k \right]  \label{eq:fidSYM}\\
  & \cong & \frac{1}{d} \left[ 1 + (d - 1) \left( 1 - \frac{d}{N} \right)^k
  \right], 
\end{eqnarray}
where the approximation holds in the asymptotic limit of large number of
copies.

From the above results we can readily obtain some conclusions on how large a
system needs to be in order to be considered ``classical'' as far as the
readout of the information is concerned. The minimum size $N$ is related to
the number of independent observations we may perform on it and still get good
estimates. For parallel spins, we see that we need a minimum size of the order
\begin{equation}
  N \sim k^{\alpha} ; \hspace{1em} \alpha > 1
\end{equation}
if we wish to obtain the classical behavior
\begin{equation}
  F_k \to 1.
\end{equation}
For smaller sizes $N \sim k^{\alpha}$ with $\alpha < 1$ the fidelity
inevitably drops to that of the random-guessing strategy,
\begin{equation}
  F_k \to \frac{1}{2} .
\end{equation}
For the optimal recycling of information, we see that, for qubits,
\begin{equation}
  F_k \to 1 \hspace{1em} \text{\tmtextrm{if $N \sim k^{\alpha}$, $\alpha > 1 /
  2$}},
\end{equation}
hence, in this case we just need a size square root of the number of
observations for a system of spins to be considered classical.

Note that the above is not in contradiction with the result
{\cite{boileau:032105}} where the authors obtain $k = O (N^2)$ for what we
call symmetric encoding into parallel spins. The quantity considered in
{\cite{boileau:032105}}, related to longevity, $k$, of a directional reference
carried by a quantum system, is the first moment of the spin-projection
operator for the state after $k$ uses, $\left\langle J_{\tmmathbf{n}(\psi)}
\right\rangle_{\rho_k (\psi)} = (2 F_k - 1) (N + 2) / 2$, which they require
to stay above arbitrary but {\tmem{fixed}} threshold $c$. We require that the
threshold approaches $(N + 2) / 2$ for all $N$ and we take the limit $N
\rightarrow \infty$.

\section{Weak measurements}\label{sec:WeakMeasurements}

We aim now to generalize the problem to situations where the
observers do not pursue the mere maximization of their estimation fidelity,
but adopt strategies where the information on the unknown state is
redistributed in different ways among the various independent observers. In
the following sections we will study the case where $K$ observers estimate the original state
with equal, but maximal, fidelity (egalitarian strategy) and the case where
all observers use the same measurement apparatus and the goal is to optimize
the estimation fidelity of the $k$th observer.

In both instances the measurements performed need to be weak, i.e. in general not
extracting all of the extractable information and hence 
inflicting less disturbance to the state. As in the greedy-observers scenario, it
suffices to consider covariant measurements. All $U$-covariant POVMs (with
outcomes labeled by the guesses) have operator density of the form
\begin{eqnarray}
  & \tilde{\mathcal{M}} (\psi) \sim U \left( g \right) S_{\tmop{ref}} U
  \left( g \right)^{\dagger}, &  \label{eq:GeneralCovariantPOVM}
\end{eqnarray}
with
\begin{eqnarray}
  & \psi = g \psi_{\tmop{ref}} g^{\dagger}, U \left( g \right) = g^{\otimes
  N}, g \in S U (d), &  \label{eq:GeneralCovariantPOVMPsi}
\end{eqnarray}
where $S_{\tmop{ref}}$ can be a positive operator commuting with $\{U \left( g
\right) ; g \in G_{\tmop{ref}} \}$, where $G_{\tmop{ref}} \subset S U (d)$ is
the set of unitaries which leave the reference state $\psi_{\tmop{ref}}$
invariant {\cite{Holevo1982Prob.andStat.Asp.of.Q.Theory}} {\emdash} the
completeness POVM relation can be easily imposed in this covariant
construction.

It is clear that, for optimal weak measurements, the post-measurement states
will not in general be pure anymore. They will depend not only on the
measurement outcome (guess) of the current observer but on particular guesses
of all predecessing observers and the preparation parameter $\psi_0$. The
probability density of obtaining measurement outcome leading to a guess
$\psi_k$, given the previous observer has obtained the guess $\psi_{k - 1}$,
is not independent of previous observers' guesses, i.e. $\tilde{p}_k (\psi_k |
\psi_{k - 1}, \ldots, \psi_0) = \tilde{p} (\psi_k | \psi_{k - 1})$ does not
hold in general. Thus, we have to start with the histories decomposition
Eq.~(\ref{eq:p(k|0)ChannelApproachEverythingCovariant}) since
Eq.~(\ref{eq:p(k|0)EverythingCovariant}) does not simplify to
Eq.~(\ref{eq:p(k|0)IntoHistoriesFullDisturb}) anymore.

To proceed further, we will hence follow the approach outlined in
Eq.~(\ref{eq:p(k|0)ChannelApproachEverythingCovariant}), where to the $k$th
observer, the action of all previous observers is described as covariant
channels. We hence need to calculate actions of the channels $\chi_k, k = 1,
\ldots, K - 1$. We will do that in what follows for the single qu$d$it case
and for the qubit case restricted to encoding into copies.

\subsection{Single
copy: arbitrary dimension}\label{sec:EqualitarianSingleQudit}

We start with the case of a (single copy) of an unknown pure state $\psi_0 =
\left| \psi_0 \rangle \langle \psi_0 \right|$ of arbitrary finite dimension
$d$. A qu$d$it being measured using a $S U (d)$-covariant instrument
undergoes, if the measurement outcome is unknown, the dynamics given by the
channel $\chi_k$ which is $S U (d)$-covariant, i.e. a convex combination of
the identity channel and the contraction to the total mixture, acting as
\begin{equation}
  \label{eq:InvariantQuditChannel} \chi_k ( \hat{\rho}) = r_k \hat{\rho} + (1
  - r_k) \mathbbm{1} / d.
\end{equation}
On the other hand, the $k$th observer's fidelity of the guess of an original
reference state $\psi_{\tmop{ref}}$, for an effectively encoded state
$\hat{\rho}_{\tmop{ref}}^{(k - 1)} = \chi_{k - 1} \circ \ldots \circ \chi_1 (
\left| \psi_{\tmop{ref}} \rangle \langle \psi_{\tmop{ref}} \right|)$ -- the
result of sending $| \psi_{\tmop{ref}} \rangle \langle \psi_{\tmop{ref}} |$
through the $S U (d)$-covariant channels $\chi_1, \ldots, \chi_{k - 1}$ -- is
given by
\begin{eqnarray}
  F_k & = & \sum_o \int \mathd \mu \left( g \right) \tmop{Tr} \left[ g \left.
  | \psi_{\tmop{ref}} \right\rangle \langle \psi_{\tmop{ref}} |g^{\dagger} g_o
  | \psi_{\tmop{ref}} \rangle \left\langle \psi_{\tmop{ref}} \right|
  g^{\dagger}_o \right] \nonumber\\
  &  & \times \tmop{Tr} \left[ g \hat{\rho}_{\tmop{ref}}^{(k - 1)}
  \hspace{0.25em} g^{\dagger} M^{(k)}_o \right], 
  \label{eq:FkUninterpretedOutcomes}
\end{eqnarray}
where the state we wish to estimate is $g \left| \psi_{\tmop{ref}} \rangle
\langle \psi_{\tmop{ref}} \right| g^{\dagger}$. For convenience, we assume the
(currently) last observer's POVM to be one with finitely many outcomes denoted
by $o$. The guess associated with an outcome is denoted by $\psi_o = g_o
\left| \psi_{\tmop{ref}} \rangle \langle \psi_{\tmop{ref}} \right|
g_o^{\dagger}$. Using Eq.~(\ref{eq:FundReprInteglal}) of
Appendix~\ref{sec:SingleKrausChannel-SU(d)} we obtain
\begin{equation}
  \label{eb20.06.07-2} F_k = \frac{(d O^{(k - 1)}_S - 1) O^{(k)}_M}{d (d + 1)
  (d - 1)} + \frac{d - O^{(k - 1)}_S}{(d + 1) (d - 1)},
\end{equation}
where $O^{(k - 1)}_S$ is the overlap of the states
\begin{equation}
  \label{eq:StateOverlap} O^{(k - 1)}_S = \tmop{Tr} \left[ \psi_{\tmop{ref}}
  \hat{\rho}_{\tmop{ref}}^{(k - 1)} \right],
\end{equation}
and $O^{(k)}_M$ is the overlap between guesses and corresponding POVM elements
\begin{equation}
  \label{eq:EffectOverlap} O^{(k)}_M = \sum_o \tmop{Tr} \left[ \psi_o
  M^{(k)}_o \right] .
\end{equation}

For a general $S U (d)$-covariant qudit channel,
Eq.~(\ref{eq:InvariantQuditChannel}), induced by the $k$th measurement and the
averaging due to lack of knowledge about it, one trivially has
\begin{eqnarray}
  F_{k + 1} & = & r_k \sum_q \int \mathd g \hspace{0.25em} \tmop{Tr} \left[ g
  \left. | \psi_0 \right\rangle \langle \psi_0 | \right. \nonumber\\
  &  & \times g^{\dagger} g_q \left| \psi_{\tmop{ref}} \right\rangle
  \left\langle \psi_{\tmop{ref}} \right| g^{\dagger}_q \text{]} \nonumber\\
  &  & \times \tmop{Tr} \left[ g \hat{\rho}_{\tmop{ref}}^{(k - 1)}
  g^{\dagger} M^{(k + 1)}_q \right] + \frac{1 - r_k}{d} \nonumber\\
  & = & r_k \left( F - \frac{1}{d} \right) + \frac{1}{d}, 
\end{eqnarray}
where $F$ is \ the average fidelity of the $(k + 1)$th observer's guess if the
measurement would have been performed on the state $\hat{\rho}^{(k - 1)}$ --
i.e. as if the $k$th observer would have not measured at all. The fidelity has
the property $1 / d \leq F \leq 1$, where $F = 1 / d$ \ corresponds to pure
guessing without actually measuring anything. It follows that in order to
maximize the possible $F_{k + 1}$ for any fixed measurement $M^{(k + 1)}$ one
has to have $r_k$ as large as possible. Naturally, $r_k$ will be ultimately
limited by the achieved $F_k$ but also by particular choice of the $k$th
observer's measurements (instrument) attaining that value of the fidelity.

At this stage we have to be more explicit in describing the observers'
measurement apparata. In particular we have to specify the instrument
realizing the POVM $M^{(k)}$ that appears in
Eqs.~(\ref{eq:FkUninterpretedOutcomes}) and (\ref{eq:EffectOverlap}). Here, we
have two options: The first option is that the quantum operation performed,
upon obtaining any outcome $o$, is given by a single-term Kraus decomposition.
That is the unnormalized post-measurement state for an outcome $o$ is given by $\hat{\rho}_{\tmop{post}} = A_o^{\dagger} \hat{\rho}_{\tmop{in}}
A_o$ (we omit the index indicating the outcome in the post-measurement states). The second option is that there exist some outcomes for which the
operation has multiple Kraus operators in its decomposition ($\exists \alpha :
\hat{\rho}_{\tmop{post}} = \sum^n_{i = 1} B_{\alpha, i}^{\dagger}
\hat{\rho}_{\tmop{in}} B_{\alpha, i} ; n > 1 ; \forall i, B_{\alpha, i} \neq
\hat{0},$). In the latter case we formally redefine the POVMs used in
Eqs.~(\ref{eq:FkUninterpretedOutcomes}) and (\ref{eq:EffectOverlap}) -- we
simply use the language of the fine-grained measurement with POVM elements
$M'_{\alpha, i} \assign B_{\alpha, i}^{\dagger} B_{\alpha, i}$ and operations
defined by $\hat{\rho}_{\tmop{post}} = B_{\alpha, i}^{\dagger}
\hat{\rho}_{\tmop{in}} B_{\alpha, i}$ for all $\alpha$s where a multi-term
Kraus decomposition would otherwise take place. Since the additional labels
$i$ are not used for anything (they are not accessible to the observer and
thus can't influence his guess), these new formal apparata provide an
equivalent description. Thus, we can always assume a description of the
measurement process in terms of an apparatus with single-term Kraus
decomposition for each outcome. For such apparatus, the averaged effective
channel is of the form \ Eq.~(\ref{eq:InvariantQuditChannel}), with the
parameter $r_k$ acquiring a particularly simple form (see
Appendix~\ref{sec:SingleKrausChannel-SU(d)})
\begin{equation}
  \label{eq:cDefMain} \text{$r_k = \frac{c - 1}{(d + 1) (d - 1)}$, where \ $c
  = \sum_{o } \left| \tmop{Tr} A^{\left( k \right)}_o \right|^2$} .
\end{equation}
Recall that we wish to have $r_k$ as large as possible given the $k$th
observer's achieved fidelity $F_k$, i.e., in the language of the
single-Kraus-term apparata, the value of $c$ as large as possible.

For a given value of $F_k$, a measurement both reaching $F_k$, i.e. the
required $O^{(k)}_M$ in Eq. (\ref{eb20.06.07-2}), and maximizing $c$ in
Eq.~(\ref{eq:cDefMain}), is known to be given by {\cite{PhysRevLett.86.1366}}
\begin{equation}
  \label{eq:QuditWeakKrausOperator} A^{(k)}_a = \sqrt{\frac{O^{(k)}_M}{d}} |a
  \rangle \langle a| + \sqrt{\frac{d - O^{(k)}_M}{d (d - 1)}} \left(
  \mathbbm{1} - |a \rangle \left\langle a| \right) \right.,
\end{equation}
where $\{ \left| a \right\rangle \}_{a = 1}^d$ is an arbitrary orthonormal
basis. Thus the largest $c$, given $F_k$ (i.e. given $O^{(k)}_M$), is
\begin{equation}
  \label{eq:copt} c = \left[ \sqrt{O^{(k)}_M} + \sqrt{(d - 1) (d - O^{(k)}_M)}
  \right]^2 .
\end{equation}
The corresponding POVM reads
\begin{eqnarray}
  M^{(k)}_a & = & A^{(k) \dagger}_a A^{(k)}_a \nonumber\\
  & = & \frac{O^{(k)}_M - 1}{d - 1} |a \rangle \langle a| + \frac{d -
  O^{(k)}_M}{d (d - 1)} \mathbbm{1}, 
\end{eqnarray}
i.e. for this particular POVM the optimal instrument in terms of Kraus
operators is given by the Hermitian square-root $A^{(k)}_a =
\sqrt{M^{(k)}_a}$. One could continue the analysis using the (unaveraged)
$d$-outcome measurement, Eq.~(\ref{eq:QuditWeakKrausOperator}), optimal for
any achievable value of $F_k$. However, we will proceed in terms of the
effective, averaged, covariant apparata with measurement ``outcomes'' given by
the possible guesses. Covariant apparata may be easily constructed {\emdash}
this is a much harder task in the case of discrete apparata for encodings into
higher-dimensional Hilbert spaces (c.f. {\cite{PhysRevLett.81.1351}} for $N$
copies of a qubit).

It follows from Eq.~(\ref{eq:GeneralCovariantPOVM}) that any $S U
(d)$-covariant POVM on a qu$d$it (with outcomes corresponding to guesses) has
the operator density of the form
\begin{equation}
  \label{eq:CovMeasQudit} \tilde{\mathcal{M}}^{(\varepsilon_k)} (\psi_k) = (1
  - \varepsilon_k) \mathbbm{1} + \varepsilon_k \tilde{\mathcal{M}} (\psi_k),
\end{equation}
where $\tilde{\mathcal{M}}$ is the operator density of the optimal covariant
POVM of the greedy-observers scenario, Eq.~(\ref{eq:CovMeasNCopies}), and
$\varepsilon_k$ parametrizes the strength of the measurement.

Then, Eq.~(\ref{eq:CovMeasQudit}) leads to
\begin{equation}
  \label{eq:MeasurementOverlapWeak} O^{(k)}_{\mathcal{M}^{(\varepsilon_k)}} =
  1 + \varepsilon_k (d - 1),
\end{equation}
where we have used Eq.~(\ref{eq:EffectOverlap}) in the form
$O^{(k)}_{\mathcal{M}^{(\varepsilon_k)}} = \int \mathd \psi_k \tmop{Tr} \left[
\psi_k \tilde{\mathcal{M}}^{(\varepsilon_k)} \left( \psi_k \right) \right]$.
Note that $1 \leq O^{(k)}_{\mathcal{M}^{(\varepsilon_k)}} \leq d$, depending
on the ``greediness'', or strength, $\varepsilon_k$, $0 \leq \varepsilon_k
\leq 1$, of the $k$th observer's measurement. Constructing the corresponding
Hermitian-square-root Kraus operators $\mathcal{A}^{(\varepsilon_k)}$ defined
by
\begin{eqnarray}
  \tilde{\mathcal{A}}^{(\varepsilon_k)} \left( \psi \right) & = &
  \sqrt{O^{(k)}_{\mathcal{M}^{(\varepsilon_k)}}} | \psi \rangle \langle \psi |
  \nonumber\\
  &  & + \sqrt{\frac{d - O^{(k)}_{\mathcal{M}^{(\varepsilon_k)}}}{(d - 1)}}
  \left( \mathbbm{1} - | \psi \rangle \left\langle \psi | \right),
  \label{eq:QuditWeakKrausOperatorCovariant} \right.
\end{eqnarray}
($\tilde{\mathcal{A}}$ is the operator density of the Kraus
operator {\footnote{The square-root of a measure here is only a formal
notation. In expressions where probabilities and post-measurement states are
calculated, the measure always appears to the first power. A rigorous
treatment of Radon-Nikodym derivatives of quantum instruments can be found in
{\cite{Davies1976QThOfOpenSys}} and {\cite{J.Math.Phys.39.1373}}.}}
$\mathcal{A}$, i.e. $\mathcal{A} \left( \sqrt{\mathd \psi} \right) =
\sqrt{\mathd \psi} \tilde{\mathcal{A}}_{\psi}$), we may verify that for a
given value of $F_k$ it induces the same channel as the minimal optimal
measurement, Eq.~(\ref{eq:QuditWeakKrausOperator}). Thus, the
Hermitian-square-root realization of the general weak covariant POVM,
Eq.~(\ref{eq:CovMeasQudit}), gives the optimal covariant instrument.

Using Eqs.~(\ref{eq:MeasurementOverlapWeak}) and (\ref{eb20.06.07-2}) we have
\begin{equation}
  \label{eq:FkOS} F_k = \frac{1}{d} + \frac{(d \hspace{0.25em} O^{(k - 1)}_S -
  1) \varepsilon_k}{d (d + 1)}
\end{equation}
and from Eq.~(\ref{eq:InvariantQuditChannel}) with a pure initial state, Eq.
(\ref{eq:StateOverlap}) reads
\begin{equation}
  O^{(k)}_S = \frac{1}{d} + \frac{d - 1}{d} \prod_{\beta = 1}^k r_{\beta} .
\end{equation}
Substituting the above equation into the Eq.~(\ref{eq:FkOS}) we finally obtain
\begin{equation}
  \label{eq:FkWeakQudit} F_k = \frac{1}{d} + \frac{\varepsilon_k (d - 1)}{d (d
  + 1)} \prod_{\beta = 1}^{k - 1} r_{\beta} .
\end{equation}
where the \ $r_{\beta}$ \ can be expressed as function of the parameter
$\varepsilon_{\beta}$ by making use of Eqs. (\ref{eq:cDefMain}),
(\ref{eq:copt}) and (\ref{eq:MeasurementOverlapWeak}),
\begin{equation}
  \label{eq:rbeta} r_{\beta} = \frac{d - 1 + (2 - d) \varepsilon_{\beta} + 2
  \sqrt{1 + \varepsilon_{\beta} (d - 1)} \sqrt{1 - \varepsilon_{\beta}}}{d +
  1} .
\end{equation}
The above equation applies to general consecutive measurements on a
$d$-dimensional system (single copy). Naturally, here we recover the results
of the greedy scenario, Eq. (\ref{eq:fidSYM}) with $N = 1$, in the limit of
most-informative measurements ($\varepsilon_k = 1$). 

We next 
discuss the weak measurement case for $N$ copies of a qubit state.

\subsection{$N$ copies of a qubit}\label{sec:NCopiesOfQubit}

Here we consider a signal that is a state of $N$ copies of a
two-dimensional unknown pure state. Again, we will assume that the observers
measurements are restricted to the completely symmetric Hilbert space. Hence,
this situation can be mapped to the problem of estimating the state of a
single $D$-dimensional system ($D = d_N^{\tmop{sym}} = N + 1$) which is,
however, known to belong to the restricted set of states from the orbit of a
reference pure $N$-copy state generated by elements of the range of the
symmetric representation of $S U (2)$.

It turns out to be very useful for our purposes to recall an
old result by Holevo {\cite{Holevo1982Prob.andStat.Asp.of.Q.Theory}}. He
solved the unconstrained, or in our terminology greedy, optimization problem
for generic mixed states $\varrho (\psi)$ drawn from a covariant family of
states
\begin{equation}
  \label{eq:HolevoStates} \left\{ \varrho (\psi) = U (g) \varrho (\psi_{\tmop{ref}}) U
  (g)^{\dagger} \nocomma, U (g) = g^{\otimes N}, g \in S U (2) \right\},
\end{equation}
where $\forall g \in S U \left( 2 \right) : \left[ g, \psi_{\tmop{ref}} \right] = 0
\Rightarrow \left[ g^{\otimes N}, \varrho \left( \psi_{\tmop{ref}} \right) \right] = 0$.

He finds that the optimal fidelity $F (\varrho)$ is given by
\begin{equation}
  \label{eq:fidgridnc} F_{} = \frac{1}{2} \left( 1 + \frac{2 \left\langle
  J_{\tmmathbf{n} (\psi_{\tmop{ref}})} \right\rangle_{\varrho (\psi_{\tmop{ref}})}}{N + 2} \right),
\end{equation}
where
\begin{equation}
  \left\langle J_{\tmmathbf{n} (\psi)} \right\rangle_{\varrho (\psi)} \assign
  \tmop{Tr} \left[ J_{\tmmathbf{n} (\psi)} \varrho (\psi) \right],
\end{equation}
and $J_{\tmmathbf{n} (\psi)}$ is \ the angular moment component in the
direction fixed by the Bloch vector $\tmmathbf{n} (\psi)$. The optimal greedy
covariant POVM is also proven to be given by $\tilde{\mathcal{M}}^{} (\psi) =
d^{\text{\tmtextrm{sym}}}_N \left| \psi \rangle \langle \psi \right|^{\otimes
N} \nocomma$ independently of a particular family Eq.~(\ref{eq:HolevoStates}).

The most general covariant POVMs one should consider here is one which
has a seed that commutes with $J_{\tmmathbf{n} (\psi_{\tmop{ref}})}$
{\cite{Holevo1982Prob.andStat.Asp.of.Q.Theory}}, i.e., the seed is diagonal in the
$\text{$\{ \left| j m \right\rangle \}$ }$basis (where we have chosen \
$\tmmathbf{n} (\psi_{\tmop{ref}})$ as our quantization axis). Note that, in principle, several
weakness parameters could be included in the optimization, in contrast to the
single parameter required in the previous subsection.  Here we will make the 
simplifying assumption of considering single parameter families of POVMs. That is, the  
measurement is given by a covariant POVM that, as before, is a convex combination
of the optimal greedy POVM and the full identity, i.e.
\begin{equation}
  \label{eq:CovPOVMWeakRestricted} \tilde{\mathcal{M}}^{(k), \varepsilon_k}
  (\psi) = (1 - \varepsilon_k) \mathbbm{1} + \varepsilon_k
  \tilde{\mathcal{M}}^{(k)} (\psi),
\end{equation}
where $\tilde{\mathcal{M}}^{(k)}$ given above, \ $0 \leq \varepsilon_k \leq 1$
parametrizing the strength of the $k$th observer's measurement, and (ii) with
the corresponding Hermitian-square-root Kraus operators densities:
\begin{eqnarray}
  \tilde{\mathcal{A}}^{\left( k \right), \varepsilon_k} (\psi) & = & b_k
  \frac{\tilde{\mathcal{M}}^{(k)} (\psi)}{d^{\tmop{sym}}_N} + a_k \mathbbm{1},
  \label{eq:NQuditWeakKrausOperator}
\end{eqnarray}
where
\begin{eqnarray}
  a_k & = & \sqrt{1 - \varepsilon_k},  \label{eq:ak}\\
  b_k & = & \sqrt{1 + (d^{\tmop{sym}}_N - 1) \varepsilon_k} - a_k . 
  \label{eq:bk}
\end{eqnarray}

It is shown in Appendix~\ref{app:channel} that such evolution leads to a
channel that leaves the post-measurement state, after averaging over guesses,
diagonal in the $\{ \left| j m \right\rangle \}$ basis, and hence it is of the
form Eq.~(\ref{eq:HolevoStates}). We emphasize that although the above two
restrictions seem to be a reasonable guess for a generalization of the optimal
apparatus from the single-copy case, we do not have a proof that, for $N > 1$,
such apparata really are among the optimal ones. Therefore, it is only
guaranteed that we obtain a lower bound $\mathbbm{F}_{\tmop{eq}}$ on the
maximum $\mathcal{F}_{\tmop{eq}}$, i.e. $\mathbbm{F}_k \leq
\mathcal{F}_{\tmop{eq}}$.

We start by rewriting the average fidelity
$\mathcal{\mathbbm{F}}^{\varepsilon_k}_1$ of single estimation using the
($\varepsilon_k$-strong) apparatus, Eq.~(\ref{eq:CovPOVMWeakRestricted}), of
the $k$th observer's measuring on arbitrary set of states in terms of
estimation fidelity, $F_1$, obtained using the apparatus of the greedy
observers problem,
\begin{equation}
  \label{eq:NQubitCopiesFweak(F)} \mathbbm{F}^{\varepsilon_k}_1 = \frac{(1 -
  \varepsilon_k)}{2} + \varepsilon_k F_1
\end{equation}
Then by Eq.~(\ref{eq:fidgridnc}), for an arbitrary family of states
Eq.~(\ref{eq:HolevoStates}),
\begin{eqnarray}
  \mathbbm{F}^{\varepsilon_k}_1 & \overset{}{=} & \frac{1}{2} \left( 1 +
  \varepsilon_k \frac{2 \left\langle J_{\tmmathbf{n} (\psi)}
  \right\rangle_{\varrho (\psi)}}{N + 2} \right) . 
  \label{eq:NQubitCopiesF1weak}
\end{eqnarray}

Appendix~\ref{app:channel} gives us the post-measurement states for each step
in a sequence of weak measurements given by
Eq.~(\ref{eq:NQuditWeakKrausOperator}) and thus, for each $k$, we can evaluate
the average fidelity of the $k$th observer $F^{\tmmathbf{\varepsilon}}_k$
where $\tmmathbf{\varepsilon} = (\varepsilon_k, \ldots, \varepsilon_1)$.
Formally this is done according to Eq.~(\ref{eq:NQubitCopiesF1weak}) with
$\varrho (\psi) \mapsto \hat{\rho}_{k - 1}^{\tmmathbf{\varepsilon}} =
\chi^{\varepsilon_{k - 1}} \circ \ldots \circ \chi^{\varepsilon_1} (\varrho
(\psi))$, i.e.
\begin{equation}
  \label{eq:Fidelita-k-teho-weak} \mathbbm{F}^{\tmmathbf{\varepsilon}}_k
  =\mathbbm{F}^{\varepsilon_k}_1 ( \hat{\rho}^{\tmmathbf{\varepsilon}}_{k -
  1}) = \frac{1}{2} \left( 1 + \varepsilon_k \frac{2 \left\langle
  J_{\tmmathbf{n}} \right\rangle_{\hat{\rho}^{\tmmathbf{\varepsilon}}_{k -
  1}}}{N + 2} \right),
\end{equation}
where the following relation holds between the consecutive output states
(Appendix~\ref{app:channel}),
\begin{equation}
  \label{eq:SpinExpValueEvolution-weakMAin} \left\langle J_z \right\rangle_{k
  + 1} = \left( a_{k + 1}^2 + \frac{2 a_{k + 1} b_{k + 1}}{2 j + 1} +
  \frac{b_{k + 1}^2 j}{\left( j + 1 \right) \left( 2 j + 1 \right)} \right)
  \left\langle J_z \right\rangle_k .
\end{equation}
We are now in position to calculate two interesting scenarios where weak
measurements need to be considered.

\section{Egalitarian strategy }\label{sec:FairObservers}

We devise a protocol such that the estimation fidelity 
obtained by each observer is the same and maximal, i.e. we want to find the
maximum fidelity, $F_k \equiv \mathcal{F}_{\tmop{eq}}$, under the egalitarian
constraints $F_k = F_1$, $\forall k \in \{2, \ldots, K\}$. The overall number
of observers, $K$, is fixed beforehand and each observer knows his tally
number $k$ in the sequence. One can visualize this scenario as different apparata
being delivered to the observers by an external party, or as observers
sharing a single apparatus which adjusts its measurement-strength automatically
before each measurement. We again require that each observer orients his
apparatus independently, and do not allow communication between observers.

With these conditions, it is clear that the last observer will perform an
optimal greedy measurement for the ensemble of states on the input of his
apparatus, while going backwards each of his predecessor's measurement will be
weaker and weaker, i.e. less and less demolishing.

With these considerations in mind we can give the results for the two types
of encodings discussed in Subsections~\ref{sec:EqualitarianSingleQudit} and
\ref{sec:NCopiesOfQubit}.

\subsection{System of arbitrary dimension (single
copy)}\label{sec:EgalQudit}

Using Eq.~(\ref{eq:FkWeakQudit}) the condition $F_k = F_{k - 1}$ translates
into
\begin{equation}
  \label{eq:RecursionRelationQuditImplicit} \varepsilon_{k - 1} =
  \varepsilon_k r_{k - 1},
\end{equation}
or, more explicitly through Eq.~(\ref{eq:rbeta}),
\begin{equation}
  \label{eq:RecursionRelationQudit} \varepsilon_{k + 1} = \frac{\varepsilon_k
  (d + 1)}{d - 1 + (2 - d) \varepsilon_k + 2 \sqrt{1 + \varepsilon_k (d - 1)}
  \sqrt{1 - \varepsilon_k}},
\end{equation}
where the initial condition $\varepsilon_K = 1$ follows from the fact that, as
mentioned above, the last, $K$th, observer can measure greedily as there is no
subsequent observer to care about. The recursion relations
Eq.~(\ref{eq:RecursionRelationQudit}) are quadratic and hence can be inverted
analytically, providing all the measurement strengths $\varepsilon_k$ starting
from $\varepsilon_K = 1$. However, a closed-form solution for $\varepsilon_1$
seems to be hard to obtain for finite $K$. Nevertheless, for large $K$ we can
obtain an asymptotic analytical expression of the fidelity and the initial
$\varepsilon_k$'s.

If $K \gg 1$ we expect the first measurements to be very weak, i.e. with
$\varepsilon_k \ll 1$. Performing a Taylor expansion around $\varepsilon_k =
0$ in the recursion relation Eq.~(\ref{eq:RecursionRelationQudit}) we obtain
an approximated relation for small values of $k$

\begin{equation}
  \label{eq:RecursionRelationQuditSimplified} \varepsilon_{k + 1} =
  \varepsilon_k + \frac{d^2}{4 \left( d + 1 \right)} \varepsilon_k^3,
\end{equation}
or, by defining $\alpha(j) = \varepsilon_{K + 1 - j}$,
\begin{equation}
  \label{eq:RecursionRelationQuditSimplified-a} \alpha(j) = \alpha(j + 1) + \frac{d^2}{4
  \left( d + 1 \right)} \alpha(j + 1)^3,
\end{equation}
which holds for large values of $j$. \ In this regime $\alpha(j)$ is vanishing
small and the difference equation can be written as an ordinary differential
equation
\begin{equation}
  \label{eq:difeqaQDIT} \frac{\mathd \alpha (x)}{\mathd x} = - \frac{d^2}{4 \left(
  d + 1 \right)} \alpha \left( x \right)_{}^3
\end{equation}
which yields,
\begin{eqnarray}
  \alpha \left( j \right) & = & \sqrt{\frac{2}{\left( j - j_0 \right) d^2 / \left(
  d + 1 \right) + 2 \alpha \left( j_0 \right)^{- 2}}} \\
  & \simeq & \sqrt{\frac{2 \left( d + 1 \right)}{j d^2}}, 
\end{eqnarray}
where $j_0$ is a fixed ($j_0 \ll K$) lower boundary of integration chosen such
that the above approximations are valid. With this we finally arrive at
\begin{eqnarray}
  \varepsilon_1 = \alpha(K) \simeq \frac{1}{d} \sqrt{\frac{2 (d + 1)}{K}}, &  & (K
  \gg 1) .  \label{eq:DifferenceEquationQuditLargeK}
\end{eqnarray}
Inserting the above into $F_1$ of Eq.~(\ref{eq:FkWeakQudit}) we have, for
large $K$, the maximal average fidelity of each egalitarian observer
\begin{equation}
  \label{eq:FRMTSingleQudit} \mathcal{F}_{\tmop{eq}} (K, d) \simeq \frac{1}{d}
  \left[ 1 + \frac{d - 1}{d} \sqrt{\frac{2}{(d + 1) K}} \right] .
\end{equation}
Let us note that a related problem of information/disturbance trade-off in
sequential weak measurements on a qudit signal has been studied in
Ref.~{\cite{1742-6596-67-1-012029}}. There, users are not considered fully
independent, in particular they share a reference frame, which allows them to
obtain an estimation fidelity that does not decrease with the number of users.

\subsection{$N$ copies of a qubit}\label{sec:EgalNqubit}

From Eq.~(\ref{eq:Fidelita-k-teho-weak}) it follows that in order for every
observer to have the same fidelity ($F_k^{\varepsilon} =
F^{\tmmathbf{\varepsilon}}_l$, $\forall l, k$ s.t. $0 < l < k \leq K$), it
must hold that
\begin{equation}
  \label{eq:SameFidelityCondition} \varepsilon_k \left\langle J_n
  \right\rangle_{\hat{\rho}^{\tmmathbf{\varepsilon}}_{k - 1}} = \varepsilon_l
  \left\langle J_{\tmmathbf{n}}
  \right\rangle_{\hat{\rho}^{\tmmathbf{\varepsilon}}_{l - 1}} .
\end{equation}
To proceed further we need to evaluate how the channels $\chi^{\varepsilon_k}$
transform $\left\langle J_{\tmmathbf{n}} \right\rangle$ for the relevant
states, which we do in Appendix~\ref{app:channel}. Comparing
Eq.~(\ref{eq:SameFidelityCondition}), with $l = k + 1$, to
Eq.~(\ref{eq:SpinExpValueEvolution-weak}) of Appendix~\ref{app:channel} we get
a recursion relation for the strength parameters $\varepsilon_k :$
\begin{widetext}
\begin{equation}
  \label{eq:RecRelationNCopies} \varepsilon_{k + 1} = \frac{(N + 1) (N + 2)
  \varepsilon_k}{(N + 1)^2 + (N - 1) (1 - 2 \varepsilon_k) + 4 \sqrt{(1 -
  \varepsilon_k) (1 + N \varepsilon_k)} - 2},
\end{equation}
\end{widetext}
where $\varepsilon_K = 1$. Again, this recursion relation gives all the
strength parameters $\varepsilon_k$ starting on reverse from $k = K$. To
obtain the fidelity one needs to solve the recurrence relation
Eq.~(\ref{eq:RecRelationNCopies}) for $k = 1$, and then use
Eq.~(\ref{eq:Fidelita-k-teho-weak}) to get
\begin{equation}
  \label{eq:FidelitaWeak} \mathbbm{F}_{\tmop{eq}} (N, K) = \frac{1}{2} \left(
  1 + \Delta_{\tmop{eq}} \right),
\end{equation}
where
\begin{equation}
  \label{eq:DeltaWeak} \Delta_{\tmop{eq}} = \varepsilon_1 (K, N) \frac{N}{N +
  2} .
\end{equation}
The presence of the square roots in Eq.~(\ref{eq:RecRelationNCopies}) prevents
the existence of a closed expression for $\varepsilon_1$. However in the
asymptotic regimes of large $K$ or large $N$ \ we can find the leading order
behaviors of $\varepsilon_1$ and $\mathcal{\mathbbm{F}}_{\tmop{eq}}$.

Let us first consider the situation $K \gg N$. As above, we expect the first
measurements to be very weak, i.e with $\varepsilon_k \ll 1$, $k \leq k_0 \ll
K$. Thus we can do a Taylor expansion \ in $\varepsilon$ around the point
$\varepsilon = 0$ in Eq.~(\ref{eq:RecRelationNCopies}) and get the approximate
relation
\begin{equation}
  \label{eq:NQubitAsymptDifference} \varepsilon_{k + 1} = \varepsilon_k +
  \frac{N + 1}{2 (N + 2)} \varepsilon_k^3,
\end{equation}
which, proceeding as in
Eqs.~(\ref{eq:RecursionRelationQuditSimplified})-(\ref{eq:DifferenceEquationQuditLargeK}),
leads to
\begin{equation}
  \label{eq:NQubitAsymptSolution} \varepsilon_1 \simeq \sqrt{\frac{N + 2}{(N +
  1) K}}, \hspace{1em}  \left( K \gg N \right) .
\end{equation}
Inserting Eq.~(\ref{eq:NQubitAsymptSolution}) into Eq.~(\ref{eq:DeltaWeak}),
we have
\begin{eqnarray}
  \Delta_{\tmop{eq}} \simeq \frac{N}{\sqrt{(N + 1) (N + 2) K}}, &  & (K \gg N)
  .  \label{fidelity-assymptotic}
\end{eqnarray}

Again we obtain a behaviour for large number of observers as $\Delta \sim 1 /
\sqrt{K} $. This \ result deserves some comments, since one would naively
expect that $\Delta$ degrades with the inverse of the number of observers,
i.e. as $\Delta \sim 1 / K$. The realization of the POVM
Eq.~(\ref{eq:CovPOVMWeakRestricted}) as an instrument given by
Hermitian-square-root Kraus operators, Eq.~(\ref{eq:NQuditWeakKrausOperator}),
is crucial to obtain this square root degradation. Had we used a more
destructive realization, we would indeed have obtained $\Delta \sim 1 / K$.
For instance, if we realize the POVM Eq.~(\ref{eq:CovPOVMWeakRestricted}) as a
stochastic measurement such that with probability $(1 - \varepsilon_k)$ the
outcome is just guessed, i.e. nothing is done to the state, and with
probability $\varepsilon_k$ the optimal greedy covariant measurement is
performed, the (relevant part of the) channel induced by such measurement is
$\chi'_{\varepsilon_k} = (1 - \varepsilon_k) \tmop{Id} + \varepsilon_k \chi$,
where $\tmop{Id}$ is the identity channel and $\chi$ is the channel induced by
the optimal greedy measurements. In this case
\begin{equation}
  \label{eq:NQubitJzEvolSuboptimal} \left\langle J_{\tmmathbf{n}}
  \right\rangle_{k + 1} = \left( 1 - \frac{\varepsilon_{k + 1}}{N / 2 + 1}
  \right) \left\langle J_{\tmmathbf{n}} \right\rangle_k .
\end{equation}
The condition Eq.~(\ref{eq:SameFidelityCondition}) then leads to the
recurrence relation that can easily be solved and gives
\begin{equation}
  \label{eq:NQubitRecRelSuboptimal} \varepsilon_k = \frac{N / 2 + 1}{N / 2 + K
  - k + 1} \rightarrow \varepsilon_1 = \frac{N + 2}{N + 2 K},
\end{equation}
which yields
\begin{equation}
  \label{Delta-3} \Delta = \frac{N}{N + 2 K} .
\end{equation}
and, clearly, for $K \gg N$ $\Delta \rightarrow N / \left( 2 K \right)$. Note
in addition that Eq.~(\ref{Delta-3}) is precisely the result that one obtains
with a strategy where each observer performs a greedy measurement on a
fraction $\tilde{N} = N / K$ of the copies. Indeed from
Eq.~(\ref{eq:FidSym1st}) with $d = 2$ one has $\Delta_{\tmop{eq}} = \tilde{N}
/ ( \tilde{N} + 2) = N / (N + 2 K)$.

Now we proceed to study the case where the number of copies is asymptotically
large, i.e. $N \gg 1$. The large-$N$ expansion in
Eq.~(\ref{eq:RecRelationNCopies}) yields
\begin{equation}
  \label{eq:NQubitsApproxRecRelationN<gg>K} \varepsilon_{k + 1} =
  \varepsilon_k + \frac{2 \varepsilon_k^2}{N},
\end{equation}
which, starting from $\varepsilon_K = 1 \nocomma$, gives at the order $1 / N$
\begin{equation}
  \label{eq:NQubitsApproxSolutionN<gg>K} \varepsilon_1 = 1 - \frac{2 \left( K
  - 1 \right)}{N} .
\end{equation}
Hence, from Eq.~(\ref{eq:DeltaWeak}) we have for $N \gg K$
\begin{eqnarray}
  \Delta_{\tmop{eq}} & \simeq & 1 - \frac{2 K {\color{blue} }}{N + 2} 
  \label{eq:DeltaNQubitsApproxSolutionN<gg>K}\\
  & \simeq & 1 - \frac{2 K {\color{blue} }}{N} . 
\end{eqnarray}
Naturally, for $K = 1$ in Eq.~(\ref{eq:DeltaNQubitsApproxSolutionN<gg>K}) we
recover the well known result of the optimal measurement of $N$ copies of a
qubit {\cite{PhysRevLett.74.1259,PhysRevA.71.062318}} \ given by
Eq.~(\ref{eq:fidSYM}), $d = 2$, $k = 1$. The efficiency of this egalitarian
strategy in the $N \gg K$ regime coincides with the stochastic strategy,
Eq.~(\ref{Delta-3}),  and the a greedy one where each observer measures only the
fraction $\tilde{N} = N / K$ of the copies.

In Fig.~\ref{fig:epsilon1(K)} we plot the observers' performance
$\Delta_{\tmop{eq}}$ obtained by a numerical evaluation of the exact
recurrence relation, Eq.~(\ref{eq:RecRelationNCopies}), as well as the
approximations of the limiting regimes discussed above. $N = 10^3$ has been
chosen to accommodate all the regimes. The stochastic strategy performance,  [Eq.~(\ref{Delta-3})],
which coincides 
with the greedy one over $N/K$ copies,  is
also plotted for reference. Notice that it gives a very good  approximation to the true values of 
the fidelity even for $K\gtrsim N$. The deviation starts to be appreciable only beyond $K\simeq 10^{4}$.

{\noindent}\tmfloat{h}{small}{figure}{\scalebox{0.666667}{\includegraphics{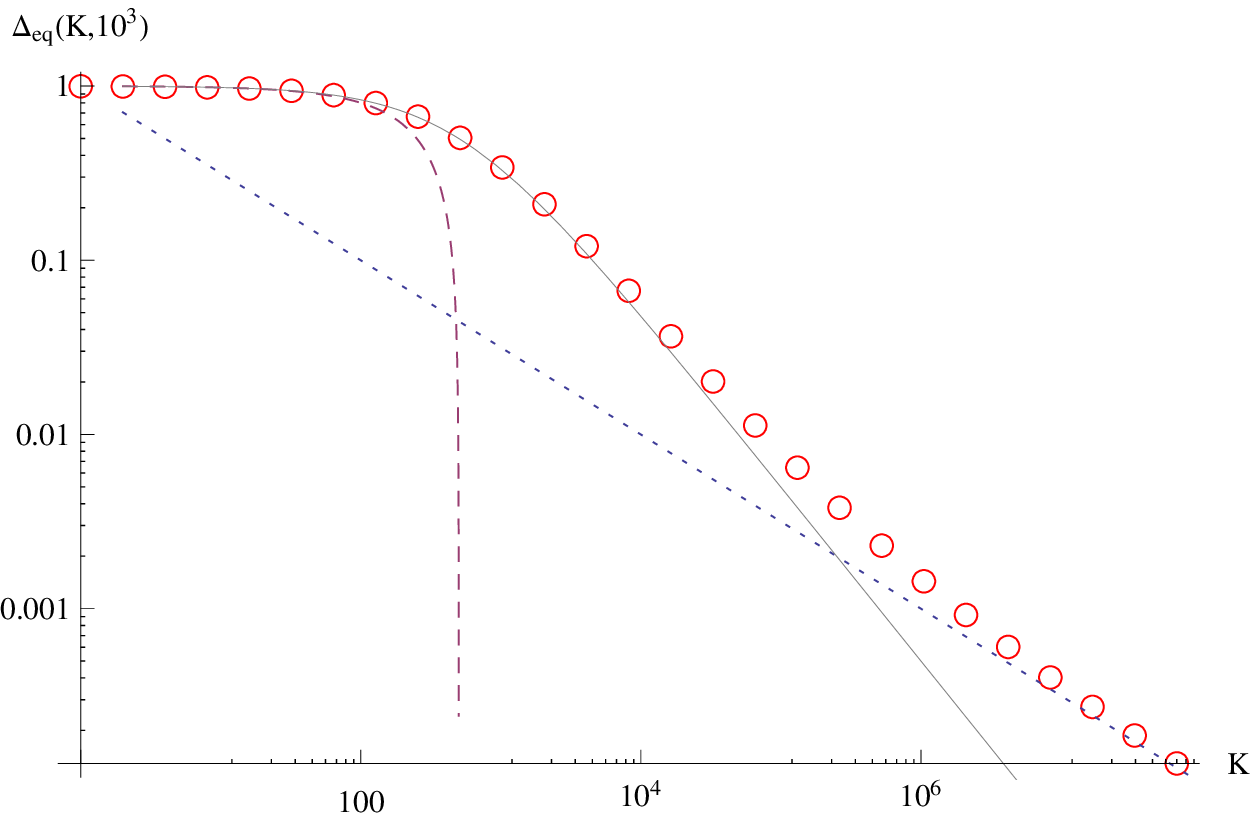}}}{\label{fig:epsilon1(K)}The
observers' measurement performance, $\Delta_{\tmop{eq}}$, as a function of the
number of observers, $K$. Solution given by the exact recurrence relations
[Eq.~(\ref{eq:RecRelationNCopies})], (circles). Approximate solution for $K
\gg N$ [Eq.~(\ref{fidelity-assymptotic})] (dotted) and $N \gg K$
[Eq.~(\ref{eq:DeltaNQubitsApproxSolutionN<gg>K})] (dashed). Stochastic
strategy [Eq.~(\ref{Delta-3})] (solid). The solid line also depicts the
performance of each observer measuring only a fraction $\tilde{N} = N / K$ of
copies.}

\section{Privileged observer strategy}\label{sec:FavorLast}

Here we consider a scenario where all observers use exactly the same
measurement device (up to the unknown relative orientation), but this is
provided, or tailored, by a particular, say the $^{} K^{}$th, observer who
wants to optimize his own estimation fidelity. That is, he has to find the
right compromise, i.e. the optimal measurement strength $\varepsilon$, \
between these two extreme cases: i) choose a very weak measurement that
prevents the $\left( K - 1 \right)$ previous observers to extract much
information from the state and thus facilitate little disturbance, but at the
same time prevents him to gain information about it when his turn comes; 2)
choose the most informative measurement that will guarantee that he extracts the
maximum information from the states he receives but, by then, all the previous
most informative measurements will have significantly ruined the input state.

\subsection{Single qudit}

For one copy,  covariant POVMs with Hermitian-square-root update rule are
optimal. They are of the form of the one-parameter
family given by Eq.~(\ref{eq:CovMeasQudit}). Based on
Eq.~(\ref{eq:FkWeakQudit}), the fidelity $F_K = (1 + (d - 1) \Delta_K) / d$ \
of the last observer is determined by
\begin{equation}
  \label{eq:QuditAllSameDeltaExact} \Delta_K = \frac{\varepsilon}{d + 1} r^{K
  - 1},
\end{equation}
where $r$ is defined in Eq. (\ref{eq:rbeta}). 

We can obtain analytical results in the asymptotic regime, $K \gg 1$. Here we again have
$\varepsilon \ll 1$ and Taylor expanding $\Delta_K$ of
Eq.~(\ref{eq:QuditAllSameDeltaExact}) around $\varepsilon = 0$ we have
\begin{eqnarray}
  \Delta_K & = & \frac{\varepsilon}{d + 1} \left( 1 - \frac{d^2
  \varepsilon^2}{4 \left( d + 1 \right)} \right)^{K - 1} 
  \label{DELTAkQuditApprox}\\
  & \simeq & \frac{\varepsilon}{d + 1} \text{} \mathe \tmop{xp} \left[ -
  \frac{d^2 K}{4 \left( d + 1 \right)} \varepsilon^2 \right] . 
  \label{DELTAkQuditMoreApprox}
\end{eqnarray}
The value of $\varepsilon^{\ast}$ that maximizes the above expression is
\begin{equation}
  \varepsilon^{\ast} = \sqrt{\frac{2 \left( d + 1 \right)}{2 d^2 K}},
\end{equation}
which inserted in Eq.~(\ref{DELTAkQuditMoreApprox}) yields
\begin{eqnarray}
  \Delta_{K, \max} \simeq \sqrt{\frac{2}{e \left( d + 1 \right) d^2 K}} &  &
  (K \gg 1)  \label{eq:QuditAllSameDeltakMax}
\end{eqnarray}
Observe that it exhibits the same characteristic square-root decay, $1 /
\sqrt{K} $, as in the egalitarian case.

\subsection{$N$ copies of a qubit}

As in the egalitarian case, we restrict our attention to weak measurements
of the type Eq.~(\ref{eq:CovPOVMWeakRestricted}) and the Hermitian-square-root
update rule. From the results of previous sections the computation of the
fidelity and measurement strength are quite straightforward. Based on
Eq.~(\ref{eq:SpinExpValueEvolution-weakMAin}) the fidelity, $F_K = (1 +
\Delta_K) / 2$, \ of the priviledges observer $K$ is determined by
\begin{eqnarray}
  \Delta_K & = & \varepsilon \frac{N / 2}{N / 2 + 1} \left\langle
  J_{\tmmathbf{n}} \right\rangle_{\rho_{k - 1}} \\
  & = & \varepsilon \frac{N}{N + 2} \left( \frac{A (\varepsilon)}{N + 1}
  \right)^{K - 1},  \label{eq:NQubitsAllSameDeltaExact}
\end{eqnarray}
where
\[ A (\varepsilon) = 2 a b + (N + 1) a^2 + \frac{N}{N + 2} b^2, \]
with $a$ and $b$ defined as in Eqs.~(\ref{eq:ak}) and (\ref{eq:bk}).

Let us obtain analytical expressions of the fidelity in the asymptotic regimes.
If  $K \gg N$,  we expect $\varepsilon \ll 1$. and Taylor expanding $\Delta_K$
around $\varepsilon = 0$ and taking two lowest orders in $\varepsilon$ we get
\begin{eqnarray}
  \Delta_K & \simeq & \frac{\varepsilon}{N + 2} \left( 1 - \frac{\left( N + 1
  \right) \varepsilon^2}{2 \left( N + 2 \right)} \right)^{K - 1} \\
  & \simeq & \frac{\varepsilon}{N + 2} \text{} \mathe \tmop{xp} \left[ -
  \frac{\left( N + 1 \right) K}{2 \left( N + 2 \right)} \varepsilon^2 \right]
  . 
\end{eqnarray}
Proceeding as in the previous subsection the optimal value of $\Delta_K$ reads
\begin{equation}
  \label{eq:NQubitsAllSameDeltaApprox2} \Delta_{K, \max} \simeq
  \sqrt{\frac{N^2}{e \left( N + 1 \right) \left( N + 2 \right) K}} .
\end{equation}
Again the fidelity degrades as $1 / \sqrt{K}$ instead of the naive $1 / K$
behavior.

In the other regime $N \gg K$ we expect
$\varepsilon \rightarrow 1$. Then, we Taylor expand
Eq.~(\ref{eq:NQubitsAllSameDeltaExact}) in the variable $(1 - \varepsilon)$
around $0$ and take terms up to the first power of $(1 - \varepsilon)$.
Maximization of $\Delta_K$ gives the optimal $\varepsilon$ which, up to the
first non-vanishing order, reads
\begin{equation}
  \label{eq:EpsilonOpt} \varepsilon = 1 - \frac{4 (K - 1)^2}{N^3} .
\end{equation}
This \ value can be taken to be \ $\varepsilon = 1 \nocomma$, as the
corrections will not affect the $1 / N \tmop{term} \tmop{of} \tmop{the}
\tmop{fidelity}$. Therefore, we should obtain the same results for the
fidelity of a greedy scenario with an asymptotically large number of copies as
discussed at the end of Section~\ref{sec:greedy}. Indeed the expansion of
$\Delta_K $ for $N \rightarrow \infty$ at the first order does not include any
\ \ $(1 - \varepsilon)$ terms and we obtain
\begin{equation}
  \Delta_{K, \max} = 1 - \frac{2 K}{N},
\end{equation}
or, equivalently,
\begin{equation}
  F_K \simeq 1 - \frac{K}{N} .
\end{equation}
Actually, we notice that in this regime, at first order in $1 / N$, a greedy strategy of
each observer in $N / K$ copies, the egalitarian and the privileged-man scenario yield
the same accuracy.

\section{Conclusions}\label{sec:Conclusion}

We have investigated to what extent can a series of independent observers
estimate a unknown state of a $d$-dimensional system by performing consecutive
measurements over the very same system. More generally, we have studied the
case where $N$ copies of a unknown state are given, and when more general
encodings into a signal system with a larger Hilbert space are permitted. This
has allowed us to assess how large does the signal system need to be so that a
given number of observers can obtain reasonable estimates, i.e. to behave
classically with regard to the readout of the state encoded in the system.
We obtain that with the optimal encoding the size has to be at least $N\sim \sqrt{K}$.
This is a quadratic improvement over the
case of a signal  consisting of copies of the encoded state for which the size must be at least
$N\sim K$.

In addition, we have studied more general ways to distribute the (limited)
information on the unknown quantum state among different observers, still
under the constraint that they measure one right after the other. We have
studied a strategy that leads to equal fidelities for all observers
(egalitarian strategy) and a second strategy where all observers are
constrained to use the same apparatus and the goal is to maximize the
estimation fidelity of a privileged observer which is the $K$'th 
position in the measurement queue. In both scenarios weak measurements are required. 
Since the systems are measured several times with observers
trying to scavenge the information contained in them after each measurement,
the choice of the Kraus operators, i.e. the choice of the instrument
implementing a given measurement, and the tradeoff fidelity 
disturbance~{\cite{PhysRevLett.86.1366, fiurasek1,fiurasek2}}, plays a crucial role. We have seen that, for
instance, the update rule given by Hermitian-square-root Kraus operators
yields, in the asymptotic regime of large number of systems, a fidelity that
degrades as the square root of the number of observers, in contrast to a
linear degradation given by a stochastic realization of the same POVM.

Our results can also shed some light on how quantum reference frames degrade
with use. This problem was first addressed  in
Refs.~{\cite{NJP.8.58,boileau:032105}} (see also {\cite{NJP.9.156}}). There,
the authors consider a setting where one has a quantum directional reference
and a set of reservoir spin particles which are pure and are either aligned or anti-aligned
randomly with respect to the reference.  One of the goals is to
correctly identify the mutual alignment of the reference and the spin particle by making a suitably (in general
collective) measurement on the two systems. As the number of spins measured
grows, the success probability of correct identification of the
orientation drops. The rate of decrease quantifies the degradation of the
directional reference with the number of measurements performed. Our results can be extended to address separable
versions of this problem, that is, settings where one measures first the
quantum reference and then one performs reference dependent tasks like, e.g.,
the one just described. These problems are under current investigation and
some further details can be found in the dissertation {\cite{P.R.PhDThesis}}.

\section*{Acknowledgments}

This work was supported by the European Union projects Q-essence, HIP 221889,
by projects CE SAV, QUTE, meta-QUTE - IMTS NFP26240120022, APVV-0673-07, VEGA
2/0092/09, by the Spanish MEC contracts FIS2008-01236, (EB) PR2010-0367, QOIT
Consolider-Ingenio 2006-00019, and by the Catalan government, CIRIT 2009GR-0985. 

\appendix\section{Evaluation of the integral
Eq.~(\ref{eq:BlochVectorEvolution})}\label{sec:BlochVectorEvolution}

Choosing, for the sake of calculations, an arbitrary reference state $\psi_0
\in \mathcal{S} (\mathcal{H}_d)$ we can parametrize the states by elements $g
\in S U (d)$ and replace the integration over the pure states by integration
over the group $S U (d)$. The integral Eq.~(\ref{eq:BlochVectorEvolution})
becomes
\begin{equation}
  \label{eq:Integral} \int_{g \in S U (d)} \mathd \mu (g) \hspace{0.25em}
  \tmmathbf{n} (g) \tilde{p} ( \hat{g} |g),
\end{equation}
where $\hspace{0.25em} \tmmathbf{n} (g)$ is a $d$-dimensional Bloch vector
parametrizing the state $\left| \psi (g) \rangle \langle \psi (g) \right|$ and
\begin{equation}
  \label{eq:CondProbTraceFormula} \tilde{p} ( \hat{g} |g) = \tmop{Tr} [
  \tilde{\mathcal{M}} ( \hat{g}) \rho (g)],
\end{equation}
where $\mathcal{S} (\mathcal{H}_D) \ni \rho (g) =\mathcal{U} (g) \rho_0
\mathcal{U} (g)^{\dagger}$. Note that due to covariance of both the
measurement and the states $\rho (g)$ it holds that
\[ \tmop{Tr} [ \tilde{\mathcal{M}} ( \bar{g} \hat{g}) \rho ( \bar{g} g)] =
   \tmop{Tr} [\mathcal{U}' ( \bar{g}) \tilde{\mathcal{M}} (
   \hat{g})\mathcal{U}' ( \bar{g})^{\dagger} \mathcal{U}( \bar{g}) \rho
   (g)\mathcal{U}( \bar{g})^{\dagger}] . \]
For optimal covariant encoding-decoding schemes it holds that the
representations are the same, i.e. $\mathcal{U}' (g) =\mathcal{U} (g)$, hence
\begin{equation}
  \label{eq:CondProbInvariance} \tilde{p} ( \bar{g} \hat{g} | \bar{g} g) =
  \tilde{p} ( \hat{g} |g) .
\end{equation}
A $d$-dimensional system in a pure state $\psi = | \psi \rangle \langle \psi
|$ can be parametrized as
\[ \psi = \frac{1}{d} \left\{ \mathbbm{1} + \kappa_d n^a T_a \right\}, \]
where
\[ \{T_a, T_b \} = \frac{\delta_{ab}}{d} + d_{ab}^c T_c, \]
with the generators defined as half the standard Gell-Mann matrices,
\[ \kappa_d = \sqrt{2 d (d - 1)}, \]
and $n^a$ are the components of a $(d^2 - 1)$-dimensional unit vector:
$\tmmathbf{n} = (n^1, n^2, \ldots, n^{d^2 - 1})$, to which we refer as Bloch
vector. This follows from imposing on $\psi$ the conditions $\tmop{Tr} \psi =
1$ and $\tmop{Tr} \psi^2 = 1$.

Not any unit vector $\tmmathbf{n}$ is allowed. By imposing the condition $\psi
= \psi^2$ we get further constrains
\begin{equation}
  \label{ebc16.10.07-1} n^a = \frac{\kappa_d}{2 (d - 2)} d_{bc}^a n^b n^c .
\end{equation}

Any state can be obtained by applying a $SU (d)$ transformation to the
reference state
\[ | \psi_0 \rangle = \left( \begin{array}{c}
     0\\
     0\\
     \vdots\\
     0\\
     1
   \end{array} \right) . \]
Note that
\[ \psi_0 = | \psi_0 \rangle \langle \psi_0 | = \frac{1}{d} \left\{
   \mathbbm{1} - \kappa_d T_{d^2 - 1} \right\}, \]
since
\[ T_{d^2 - 1} = \frac{1}{\sqrt{2 d (d - 1)}} \left( \begin{array}{ccccc}
     1 & 0 & \ldots & 0 & 0\\
     0 & 1 & \ldots & 0 & 0\\
     \vdots & \vdots & \ddots & \vdots & \vdots\\
     0 & 0 & \ldots & 1 & 0\\
     0 & 0 & \ldots & 0 & 1 - d
   \end{array} \right) \]
(the normalization ensures that $\tmop{Tr} \left[ T_{d^2 - 1}^2 \right] = 1 /
2$). Hence, the `reference' Bloch vector is
\[ \tmmathbf{n}_0 = ( \underbrace{0, 0, \ldots, 0}_{d^2 - 2}, - 1), \]
i.e., its components are
\[ n_0^{d^2 - 1} = - 1 ; \hspace{1em} n_0^a = 0 \hspace{1em} \tmop{if}
   \hspace{1em} a \neq d^2 - 1 . \]

Note that $| \psi_0 \rangle$ is invariant under $SU (d - 1) \subset SU (d)$
transformation of the form
\[ \tilde{U} \equiv U ( \tilde{g}) = \left( \begin{array}{ccccc}
     U_{1 \hspace{0.25em} 1} & U_{1 \hspace{0.25em} 2} & \ldots & U_{1
     \hspace{0.25em} d - 1} & 0\\
     U_{2 \hspace{0.25em} 1} & U_{2 \hspace{0.25em} 2} & \ldots & U_{2
     \hspace{0.25em} d - 1} & 0\\
     \vdots & \vdots & \ddots & \vdots & \vdots\\
     U_{d - 1 \hspace{0.25em} 1} & U_{d - 1 \hspace{0.25em} 2} & \ldots & U_{d
     - 1 \hspace{0.25em} d - 1} & 0\\
     0 & 0 & \ldots & 0 & 1
   \end{array} \right) . \]
Hence,
\begin{eqnarray}
  \langle \psi_0 | \psi (g) \rangle & \equiv & \langle \psi_0 |U (g) | \psi_0
  \rangle  \label{eq:SU(d-1)Invariance-Overlap}\\
  & = & \langle \psi_0 | \tilde{U} U (g) | \psi_0 \rangle = \langle \psi_0 |
  \psi ( \tilde{g} g) \rangle \nonumber\\
  & = & \langle \psi_0 |U (g) \tilde{U} | \psi_0 \rangle = \langle \psi_0 |
  \psi (g \tilde{g}) \rangle . \nonumber
\end{eqnarray}
Moreover, due to covariance of the encoding, it also has to hold that
\begin{equation}
  \label{eq:SU(d-1)Invariance-OverlapMixed} \tmop{Tr} [\rho_0 \rho (g)] =
  \tmop{Tr} [\rho_0 \rho (g \tilde{g})] .
\end{equation}
We use the group parameters $g$ to label the different states according to:
\[ \left| \psi (g) \rangle \langle \psi (g) \right| = U (g) \left| \psi_0
   \rangle \langle \psi_0 \right| U^{\dagger} (g) . \]
It follows that
\[ n^a (g) T_a = U (g) \hspace{0.25em} n^a_0 T_a \hspace{0.25em} U^{\dagger}
   (g) = A^b_a (g) n^a_0 T_b, \]
where $A^a_b ( \bar{g})$ belongs to the adjoint representation and we have
used that
\[ U (g) T_a U^{\dagger} (g) = A^b_a (g) T_b . \]
We see that
\[ n^a (g) = A^a_b (g) n^b_0 . \]
In general
\[ n^a (g) = A^a_b (g) A^b_c ( \bar{g}^{- 1}) n^c ( \bar{g}) = A^a_c (g
   \bar{g}^{- 1}) n^c ( \bar{g}), \]
from which
\[ n^a (g \bar{g}) = A^a_b (g \bar{g} \bar{g}^{- 1}) n^b ( \bar{g}) = A^a_b
   (g) n^b ( \bar{g}) . \]
Let us now consider the integral
\[ V^a ( \hat{g}) \equiv \int \mu (g) n^a (g) \tilde{p} ( \hat{g} |g) . \]
Here $\tilde{p} ( \hat{g} |g)$ is the conditional-probability density,
Eq.~(\ref{eq:CondProbTraceFormula}). Let $\bar{U} = U ( \bar{g})$ be any $S U
(d)$ transformation. We have
\begin{eqnarray*}
  V^a ( \bar{g} \hat{g}) & = & \int \mathd \mu (g) n^a (g) \tilde{p} ( \bar{g}
  \hat{g} |g)\\
  & = & \int \mathd \mu ( \bar{g}^{- 1} g) n^a ( \bar{g} \bar{g}^{- 1} g)
  \tilde{p} ( \bar{g} \hat{g} | \bar{g} \bar{g}^{- 1} g)\\
  & = & A^a_b ( \bar{g}) \int \mathd \mu ( \bar{g}^{- 1} g) n^a ( \bar{g}^{-
  1} g) \tilde{p} ( \hat{g} | \bar{g}^{- 1} g)\\
  & = & A^a_b ( \bar{g}) V^b ( \hat{g}),
\end{eqnarray*}
where we have used the invariance of the Haar measure $\mathd \mu (g)$ and the
invariance of the probability, Eq.~(\ref{eq:CondProbInvariance}).

We see that, in particular
\[ V^a ( \hat{g}) = A^a_b ( \hat{g}) V^b ( \tmmathbf{0}), \]
where $\tmmathbf{0}$ denotes the identity parameters. I.e.,
\[ V^b ( \tmmathbf{0}) = \int \mathd \mu (g) n^b (g) \tilde{p} ( \tmmathbf{0}
   |g) . \]
We now wish to show that, as expected, $V^b ( \tmmathbf{0}) \propto n_0^b$. We
proceed as follows. From
\[ T_b V^b ( \tmmathbf{0}) = \int \mathd \mu (g) T_b n^b (g) \tilde{p} (
   \tmmathbf{0} |g) \]
we observe that
\begin{eqnarray*}
  \tilde{U} T_b V^b ( \tmmathbf{0}) \tilde{U}^{\dagger} & = & \int \mathd \mu
  ( \tilde{g} g) T_b n^b ( \tilde{g} g) \tilde{p} ( \tmmathbf{0} |g)\\
  & = & \int \mathd \mu ( \tilde{g} g) T_b n^b ( \tilde{g} g) \tilde{p} (
  \tmmathbf{0} | \tilde{g} g)\\
  & = & T_b V^b ( \tmmathbf{0}),
\end{eqnarray*}
where we have used Eq.~(\ref{eq:SU(d-1)Invariance-OverlapMixed}) in the form
$\tilde{p} ( \tmmathbf{0} |g) = \tilde{p} ( \tmmathbf{0} | \tilde{g} g)$.
Hence, according to Schur's lemma, $T_b V^b ( \tmmathbf{0})$ must be the
identity in the subspace corresponding to $S U (d - 1)$, i.e., proportional to
$T_{d^2 - 1}$, from where the desired result follows immediately. Note that
from this it also follows that
\[ V^a ( \hat{g}) \propto A^a_b ( \hat{g}) n_0^b = n^b ( \hat{g}) \]
or, more explicitly,
\begin{equation}
  \label{eq:Integral-result} \int \mathd \mu (g) \tmmathbf{n} (g) \tilde{p} (
  \hat{g} |g) = \Delta \hspace{0.25em} \tmmathbf{n} ( \hat{g}),
\end{equation}
where $\Delta$ is a constant.

\section{The channel induced by the measurements
Eq.~(\ref{eq:NQuditWeakKrausOperator}) on $N$ copies of a
qubit}\label{app:channel}

We compute first the action of the channel induced by the $S O (3)$-covariant
measurement of the greedy strategy over a generic state
\begin{equation}
  \label{eq:RotInvariantState} \hat{\rho} = \sum_{m = - j}^{m = j} s_m \left|
  j m \rangle \langle j m \right| .
\end{equation}
We recall that the optimal covariant measurement in this case has the operator
density
\begin{equation}
  \tilde{\mathcal{M}} (\tmmathbf{n}) = (2 j + 1) \left| j j ; \tmmathbf{n}
  \rangle \langle j j ; \tmmathbf{n} \right|,
\end{equation}
where $\left| j j ; \tmmathbf{n} \right\rangle$ is the rotated state from the
$z$ direction (defined by the diagonalization axis of $\hat{\rho}$) into the
$\tmmathbf{n}$ direction. The channel action is given by
\begin{equation}
  \label{eq:ChannelFull} \chi ( \hat{\rho}) = (2 j + 1) \sum_m s_m \int \mathd
  n| \left\langle j m | j j ; \tmmathbf{n} \right\rangle |^2 \left| j j ;
  \tmmathbf{n} \rangle \langle j j ; \tmmathbf{n} \right|
\end{equation}
It is easy to see that this operator is invariant under rotations along the
$z$ axis and therefore is diagonal in the $\left| j m \right\rangle$ basis:
\begin{equation}
  \chi ( \hat{\rho}) = \sum_{m'} c_{m'} \left| j m' \rangle \langle j m'
  \right| .
\end{equation}
We further notice that
\begin{equation}
  \int \mathd \tmmathbf{n} \left| j j ; \tmmathbf{n} \rangle \langle j j ;
  \tmmathbf{n} \right| \otimes \left| j j ; \tmmathbf{n} \rangle \langle j j ;
  \tmmathbf{n} \right| = \frac{\mathbbm{1}^{\left( 2 j \right)}}{4 j + 1},
\end{equation}
where $\mathbbm{1}^{\left( 2 j \right)} = \sum_M \left| 2 j ; M \rangle
\langle 2 j ; M \right|$ is the projector onto the symmetric space of
dimension $4 j + 1$. Hence we have
\begin{equation}
  c_{m'} = \sum_m \Lambda^m_{m'} s_m,
\end{equation}
with
\begin{eqnarray*}
  \Lambda^m_{m'} & = & \frac{2 j + 1}{4 j + 1} | \left\langle jm, jm' | 2 j
  \hspace{0.75em} m + m' \right\rangle |^2\\
  & = & \frac{2 j + 1}{4 j + 1} \binom{2 j}{j + m} \binom{2 j}{j + m'}
  \binom{4 j}{2 j + m + m'}^{- 1},
\end{eqnarray*}
where $\left\langle j \hspace{0.25em} m, j \hspace{0.25em} m' | 2 j
\hspace{0.75em} m + m' \right\rangle$ is the Clebsch-Gordan coefficient of the
composition $\tmmathbf{} \mathbf{j} \otimes \mathbf{j} \to \mathbf{2 j}$.

As shown in the main text, to compute the fidelity it is sufficient to
calculate the expectation value of the spin component $J_z$. If the
Eq.~(\ref{eq:RotInvariantState}) is the state after $k$ uses of the channel
and $\chi ( \hat{\rho})$ the state after $k + 1$ uses, it is straightforward
to obtain
\begin{eqnarray}
  \left\langle J_z \right\rangle_{k + 1} & = & \sum_{mm'} m' \Lambda_{m'}^m
  s_m = \sum_m \frac{j}{j + 1} ms_m \nonumber\\
  & = & \frac{j}{j + 1} \left\langle J_z \right\rangle_k . 
  \label{eq:ConservationOfSpinOrientation}
\end{eqnarray}
We can now proceed to compute the action of the channel and the values of the
fidelities for the weak measurements considered in the main text. The
covariant POVM elements in this case are given by operator density
\begin{equation}
  \widetilde{\mathcal{M}} (\tmmathbf{n}) = (1 - \varepsilon) \mathbbm{1} +
  \varepsilon (2 j + 1) \left| j j ; \tmmathbf{n} \rangle \langle j j ;
  \tmmathbf{n} \right|,
\end{equation}
where the parameter $\varepsilon$ quantifies the strength of the measurement.
The corresponding Kraus operator densities are $\widetilde{\mathcal{A}}
(\tmmathbf{n}) = \sqrt{\widetilde{\mathcal{M}} (\tmmathbf{n})}$, which
explicitly read
\begin{equation}
  \widetilde{\mathcal{A}} (\tmmathbf{n}) = a \mathbbm{1} + b \left| j j ;
  \tmmathbf{n} \rangle \langle j j ; \tmmathbf{n} \right|,
\end{equation}
where $a = \sqrt{1 - \varepsilon}$ and $b = \sqrt{1 + 2 j \hspace{0.25em}
\varepsilon} - \sqrt{1 - \varepsilon}$ . The action of the channel is fully
determined from
\begin{eqnarray}
  \chi^{\varepsilon} \left( \left| j m \rangle \langle j m \right| \right) & =
  & a^2 \left| j m \rangle \langle j m \right| \nonumber\\
  &  & + a b \int \mathd \tmmathbf{n} \left\langle jm | jj, \tmmathbf{n}
  \right\rangle \nonumber\\
  &  & \times ( \left| jm \rangle \langle jj ; \tmmathbf{n} \right| + \left|
  jj ; \tmmathbf{n} \rangle \langle jm \right|) \\
  &  & + b^2 \int \mathd \tmmathbf{n}| \left\langle jm | jj ; \tmmathbf{n}
  \right\rangle |^2 \left| j j ; \tmmathbf{n} \rangle \langle j j ;
  \tmmathbf{n} \right| . \nonumber
\end{eqnarray}
Using the same techniques as in the previous case we obtain
\begin{eqnarray*}
  \chi^{\varepsilon} ( \hat{\rho}) & \equiv & \sum_{mm'} s_m
  \tilde{\Lambda}^m_{m'} s_m\\
  & = & \sum_m s_m \frac{a^2 (2 j + 1) + 2 ab}{2 j + 1} \left| j m \rangle
  \langle j m \right| + \frac{b^2}{2 j + 1} \chi ( \hat{\rho}),
\end{eqnarray*}
where $\chi ( \hat{\rho})$ is the action of the greedy channel
Eq.~(\ref{eq:ChannelFull}) and $\tilde{\Lambda}^m_{m'} = \frac{a^2 (2 j + 1) +
2 ab}{2 j + 1} \delta^m_{m'} + \frac{b^2}{2 j + 1} \Lambda^m_{m'}$.

We finally compute the relation of the expectation values of the operator
$J_z$ before and after the use of the channel. The analogue of
Eq.~(\ref{eq:ConservationOfSpinOrientation}) now reads
\begin{eqnarray}
  \left\langle J_z \right\rangle_{k + 1} & = & \sum_{mm'} m'
  \tilde{\Lambda}_{m'}^m s_m \nonumber\\
  & = & \left( a_{k + 1}^2 + \frac{2 a_{k + 1} b_{k + 1}}{2 j + 1} +
  \frac{b_{k + 1}^2 j}{\left( j + 1 \right) \left( 2 j + 1 \right)} \right)
  \nonumber\\
  &  & \times \left\langle J_z \right\rangle_k, 
  \label{eq:SpinExpValueEvolution-weak}
\end{eqnarray}
where the label $k$ in the parameters $a_k$ and $b_k$ simply take into account
that the strength $\varepsilon_k$ can vary from one measurement to another.

\section{The average channel induced by single-Kraus-operator measurements on
a single qu$d$it}\label{sec:SingleKrausChannel-SU(d)}

We  show that the optimal weak instrument, i.e. one maximizing next
observer's fidelity given current observer's fidelity, for a qu$d$it induces a
channel which has the effect of adding a portion of total mixture to the encoding state. We first collect
some mathematical results concerning unitary group integrals that will be
extensively used below. For matrices $g$ belonging to the fundamental
representation of $S U (d)$ and denoting by $\mathd \mu \left( g \right)$ the
corresponding Haar measure, we have
\[ \int \mathd \mu \left( g \right) \hspace{0.25em} g_i^j g_r^{\dagger
   \hspace{0.25em} s} = \frac{\delta_i^s \delta_r^j}{d} \]
and, similarly,
\begin{widetext}
\begin{equation}
  \label{eq:FundReprInteglal} \int \mathd \mu \left( g \right) \hspace{0.25em}
  g_i^j g^l_k g_r^{\dagger \hspace{0.25em} s} g_t^{\dagger \hspace{0.25em} v}
  = \frac{(\delta_i^s \delta_k^v + \delta_i^v \delta_k^s) (\delta_r^j
  \delta_t^l + \delta_r^l \delta_t^j)}{2 d (d + 1)} + \frac{(\delta_i^s
  \delta_k^v - \delta_i^v \delta_k^s) (\delta_r^j \delta_t^l - \delta_r^l
  \delta_t^j)}{2 d (d - 1)} .
\end{equation}
\end{widetext}

The last result can be most easily seen by writing the integral above as
\begin{eqnarray*}
&\displaystyle
\int \mathd \mu \left( g \right)\;
\left(
\Young1223{.25}9
\oplus\kern-1.em\raisebox{-.45em}{ \Young2312{.25}9}
\right)
\otimes
\left(
\Young1223{.25}9
\oplus\kern-1.em\raisebox{-.45em}{ \Young2312{.25}9}
\right)^\dagger
& \label{young tableaux}\\[-1.em]
&&\nonumber
\end{eqnarray*}
and recalling the orthogonality relations of the irreducible representations
of unitary groups, which state that
\begin{eqnarray*}
&\displaystyle
\int \mathd \mu \left( g \right)\;
\Young1223{.25}9
\otimes\kern-1.em\raisebox{-.45em}{ \Young2312{.25}9}^\dagger
=
\int \mathd \mu \left( g \right)\kern-.5em
\raisebox{-.45em}{ \Young2312{.25}9} \kern-.25em\otimes
{\Young1223{.25}9 }^{\;\dagger}
=0,
& 
\\[1.em]
&&\nonumber
\\[-1.em]
&\displaystyle
\int \mathd \mu \left( g \right)\;
\Young1223{.25}9
\otimes\Young1223{.25}9 ^{\;\dagger}
\sim\openone_{\Young1223{.0}3};\quad
\int \mathd \mu \left( g \right)\kern-.5em
\raisebox{-.45em}{ \Young2312{.25}9} \kern-.25em\otimes
\kern-1.em\raisebox{-.45em}{ \Young2312{.25}9}^{\;\dagger}
\sim\openone_{\Young2312{.0}3}.
& 
\end{eqnarray*}As we argued in Section~\ref{sec:General}, the effective apparatus, given by
the actual one and the lack of knowledge about it, is covariant (with respect
to $S U (d)$ in this case). In terms of the Kraus operators, associated to
measurement outcomes which do {\tmem{not}} transform upon a unitary
``rotation'' of the apparatus (e.g. LEDs or  dials on a display),
this means there is a unitary freedom in the next observers' possible
knowledge of those Kraus operators for any given outcome and an average is
performed over $S U (d)$. We restrict our attention to measurements with a
single term in the Kraus decomposition for any outcome -- see discussion in
Section~\ref{sec:EqualitarianSingleQudit}.

Moreover, we assume that a given observer does not know the measurement
outcomes of the previous observers, thus no other object, except the $i$th
observer's output state, its probability, and guess, depends on his
measurement outcome. Therefore we can perform the sum over all outcomes to get
the channel induced by such measurement. Hence, one way to look at the
measurement process is via the map
\begin{eqnarray*}
  \hat{\rho} & \mapsto & \hat{\rho}' = \mathcal{\chi} ( \hat{\rho})\\
  & = & \sum_o \int \mathd \mu \left( g \right) gA_o g^{\dagger} \hat{\rho}
  \hspace{0.75em} g A_o^{\dagger} g^{\dagger},
\end{eqnarray*}
where $\{o\}$ is the set of possible outcomes of the predecessing observer's
apparatus (or the set enriched by additional outcomes so that a quantum
operation performed given any outcome $o$ has single Kraus operator in its
Kraus decomposition).

Using Eq.~(\ref{eq:FundReprInteglal}) we get
\begin{equation}
  \label{eq:SingleQuditChannel} \mathcal{\chi} ( \hat{\rho}) = \frac{c - 1}{(d
  + 1) (d - 1)} \hspace{0.25em} \hat{\rho} + \frac{d^2 - c}{(d + 1) (d - 1)}
  \frac{\mathbbm{1}}{d},
\end{equation}
where
\begin{equation}
  \label{eq:cDefinition} c = \sum_o \left| \tmop{Tr} A_o \right|^2 .
\end{equation}

\bibliographystyle{./prsty}
\bibliography{./ScavengingOfQI}

\end{document}